\documentclass[prc,letterpaper,twocolumn,showpacs,showkeys,lengthcheck,
               nofootinbib,preprintnumbers,superscriptaddress]{revtex4-1}

\usepackage[utf8]{inputenc}

\usepackage{newtxtext}
\usepackage[libertine]{newtxmath}

\usepackage{bbold}
\usepackage{bm}

\usepackage{graphicx}
\usepackage[usenames,dvipsnames]{xcolor}
\usepackage{color}
\usepackage{tikz}

\usepackage[colorlinks=true,linkcolor=blue,urlcolor=blue,citecolor=blue]{hyperref}

\usepackage{amsmath,amsfonts}
\usepackage{slashed}

\usepackage[english]{babel}
\usepackage{titlesec}
\usepackage{microtype}
\usepackage{dcolumn}

\usepackage{blindtext}

\titlespacing{\section}{4pt}{12pt plus 4pt minus 2pt}{8pt plus 2pt minus 2pt}
\titlespacing{\subsection}{0pt}{12pt plus 4pt minus 2pt}{8pt plus 2pt minus 2pt}
\titlespacing{\subsubsection}{0pt}{12pt plus 4pt minus 2pt}{0pt plus 2pt minus 2pt}

\def\D{\Delta}

\def\hs{\hspace}

\def\lf{\left}
\def\rg{\right}

\newcommand{\vect}[1]{\boldsymbol{#1}}

\makeatletter
\newcommand*\if@single[3]{%
  \setbox0\hbox{${\mathaccent"0362{#1}}^H$}%
  \setbox2\hbox{${\mathaccent"0362{\kern0pt#1}}^H$}%
  \ifdim\ht0=\ht2 #3\else #2\fi
  }
\newcommand*\rel@kern[1]{\kern#1\dimexpr\macc@kerna}
\newcommand*\widebar[1]{\@ifnextchar^{{\wide@bar{#1}{0}}}{\wide@bar{#1}{1}}}
\newcommand*\wide@bar[2]{\if@single{#1}{\wide@bar@{#1}{#2}{1}}{\wide@bar@{#1}{#2}{2}}}
\newcommand*\wide@bar@[3]{%
  \begingroup
  \def\mathaccent##1##2{%
    \if#32 \let\macc@nucleus\first@char \fi
    \setbox\z@\hbox{$\macc@style{\macc@nucleus}_{}$}%
    \setbox\tw@\hbox{$\macc@style{\macc@nucleus}{}_{}$}%
    \dimen@\wd\tw@
    \advance\dimen@-\wd\z@
    \divide\dimen@ 3
    \@tempdima\wd\tw@
    \advance\@tempdima-\scriptspace
    \divide\@tempdima 10
    \advance\dimen@-\@tempdima
    \ifdim\dimen@>\z@ \dimen@0pt\fi
    \rel@kern{0.6}\kern-\dimen@
    \if#31
      \overline{\rel@kern{-0.6}\kern\dimen@\macc@nucleus\rel@kern{0.4}\kern\dimen@}%
      \advance\dimen@0.4\dimexpr\macc@kerna
      \let\final@kern#2%
      \ifdim\dimen@<\z@ \let\final@kern1\fi
      \if\final@kern1 \kern-\dimen@\fi
    \else
      \overline{\rel@kern{-0.6}\kern\dimen@#1}%
    \fi
  }%
  \macc@depth\@ne
  \let\math@bgroup\@empty \let\math@egroup\macc@set@skewchar
  \mathsurround\z@ \frozen@everymath{\mathgroup\macc@group\relax}%
  \macc@set@skewchar\relax
  \let\mathaccentV\macc@nested@a
  \if#31
    \macc@nested@a\relax111{#1}%
  \else
    \def\gobble@till@marker##1\endmarker{}%
    \futurelet\first@char\gobble@till@marker#1\endmarker
    \ifcat\noexpand\first@char A\else
      \def\first@char{}%
    \fi
    \macc@nested@a\relax111{\first@char}%
  \fi
  \endgroup
}
\makeatother

\newcommand{\be}{\begin{equation}}
\newcommand{\ee}{\end{equation}}

\def\emdash{\nobreak\hspace{0.1em}\textemdash\nobreak\hspace{0.1em}}

\begin{document}
\title{Massive Neutron Stars with a Color Superconducting Quark Matter Core}

\author{Takehiro Tanimoto}
\email[Corresponding author:~]{t.tanimoto@star.tokai-u.jp}
\affiliation{Department of Physics, School of Science, Tokai University,
4-1-1 Kitakaname, Hiratsuka-shi, Kanagawa 259-1292, Japan}

\author{Wolfgang Bentz}
\email[]{bentz@keyaki.cc.u-tokai.ac.jp}
\affiliation{Department of Physics, School of Science, Tokai University,
4-1-1 Kitakaname, Hiratsuka-shi, Kanagawa 259-1292, Japan}

\author{Ian C. Clo\"et}
\email[]{icloet@anl.gov}
\affiliation{Physics Division, Argonne National Laboratory, Argonne, Illinois 60439, USA}

\begin{abstract}
We construct the equation of state for high density neutron star matter at zero temperature using the two-flavor Nambu--Jona-Lasinio (NJL) model as an effective theory of QCD. We build nuclear matter, quark matter, and the mixed phases from the same NJL Lagrangian, which has been used to model free and in-medium hadrons as well as nuclear systems. A focus here is to determine if the same coupling constants in the scalar diquark and vector meson channels, which give a good description of nucleon structure and nuclear matter, can also be used for the color superconducting high density quark matter phase. We find that this is possible for the scalar diquark (pairing) interaction, but the vector meson interaction has to be reduced 
so that superconducting quark matter becomes the stable phase at high densities.
We compare our equation of state with recent phenomenological parametrizations
based on generic stability conditions for neutron stars.  
We find that the maximum mass of a hybrid star, with a color superconducting quark matter core, exceeds $2.01 \pm 0.04\,M_\odot$ which is the value of the recently observed massive neutron star PSR J0348+0432. The mass-radius relation is also consistent with gravitational wave observations (GW170817).
\vskip 0.5em
\noindent \textit{PhySH}: {
        Quark model;
        Asymmetric nuclear matter;
        Nuclear matter in neutron stars.
    }
\end{abstract}

\maketitle
\section{INTRODUCTION\label{sec:intro}}
The study of strongly interacting cold matter at high baryon densities\emdash neutron star matter\emdash has been a very important and active area of research for several decades~\cite{Itoh:1970uw,Baym:1971pw,Glendenning:1997wn,Heiselberg:1999mq,Dean:2002zx,
Baym:2017whm}. This field has attracted increased attention recently however, because of the observation of massive neutron stars exceeding two solar masses~\cite{Demorest:2010bx,Antoniadis:2013pzd} and gravitational wave measurements of a binary neutron star merger event~\cite{Abbott:2018exr,TheLIGOScientific:2017qsa,Evans:2017mmy}. Among the many theoretical tools used to study dense hadronic matter are nonrelativistic potential models~\cite{Gandolfi:2011xu,Gandolfi:2013baa}, effective field theories~\cite{Hebeler:2010xb,Kaiser:2012ex}, and relativistic theories based on a mean field description of nucleons interacting via meson exchange~\cite{Muller:1995ji,Chin:1974sa}. The role of hyperons in dense hadronic matter has also been of interest~\cite{Kaplan:1986yq,Dapo:2008qv}. On the other hand, studies based on effective quark theories of QCD\emdash often using the framework provided by the Nambu--Jona-Lasinio (NJL) model~\cite{Nambu:1961tp,Nambu:1961fr,Vogl:1991qt,Hatsuda:1994pi}\emdash have focused on the existence of a possible quark matter phase at high densities~\cite{Buballa:2003qv,Shovkovy:2004me,
Alford:2007xm,Fukushima:2010bq} and the role of color superconductivity~\cite{Bailin:1983bm,Alford:1997zt,Rapp:1997zu}. In these approaches the transition between the hadronic and quark matter phases has been described by using either the Maxwell or Gibbs constructions~\cite{Glendenning:1992vb,Wu:2017xaz}, or by interpolation methods based on hadron-quark continuity~\cite{Schafer:1998ef,Macher:2004vw,Masuda:2012kf}. Finite size effects in the mixed phase caused by the surface tension of various geometrical shapes, Coulomb interactions, and
charge screening have also been studied~\cite{Heiselberg:1992dx,Endo:2005zt,Lugones:2013ema}.

The aim of this paper is to present results for the equation of state of
high density matter and the properties of neutron stars, using effective quark degrees
of freedom for the description of the hadronic as well as the quark phase.
Most of the calculations done so far on hybrid star matter\emdash hadronic matter which converts into
quark matter at high baryon densities\emdash combined a relativistic hadronic
mean field theory for nuclear matter with some version of the NJL model for
quark matter, see for example Refs.~\cite{Baym:2017whm,Contrera:2014eza,Benic:2014jia,Li:2018ltg}.
More phenomenological approaches used a parametrization of the nuclear and quark matter phases in terms of 
polytropes~\cite{Abgaryan:2018gqp} or relativistic density functionals~\cite{Kaltenborn:2017hus}. 
In Ref.~\cite{Hempel:2013tfa} an interesting attempt was made to describe a mixture of elementary hadrons and quarks by using the same Lagrangian, the mixture being determined by a scalar field which increases the hadron masses
and decreases the quark masses as the baryon density increases. An application of this
approach to neutron stars can be found in Ref.~\cite{Dexheimer:2019pay}. 

The purpose of our present work differs from the above mentioned approaches
in the following respect: We wish to investigate whether a single effective quark theory of QCD, which can describe the quark structure of free hadrons~\cite{Cloet:2005pp,Cloet:2007em,Cloet:2014rja,Hutauruk:2016sug,Ninomiya:2017ggn} as well the role of quarks in bound nucleons and nuclear systems~\cite{Cloet:2005rt,Cloet:2006bq,Cloet:2015tha}, can also produce reasonable scenarios for the hadron-quark phase transition and the properties of neutron stars.  
Our work is based on the NJL model with the proper-time regularization scheme~\cite{Hellstern:1997nv}, which incorporates important aspects of confinement in hadronic systems~\cite{Bentz:2001vc}. 
The crucial point which leads to saturation of the nuclear matter binding energy in the
mean field approximation is the scalar polarizability of the in-medium nucleons~\cite{Bentz:2001vc}, i.e., the non-linear dependence of the nucleon mass on the constituent quark mass, which
arises naturally in the Faddeev approach based on the quark-quark correlations in the
scalar and axial-vector diquark channels~\cite{Cloet:2014rja}. A key question that we will address 
in this work is whether the same strength of the scalar diquark interaction\emdash which is 
required to reproduce the nucleon mass and other quark-quark correlation effects in baryons\emdash 
can also be used as the pairing interaction in color superconducting quark matter.   

In the course of our investigation, we will formulate a few conditions for
a reasonable scenario of the hadron-quark phase transition and the resulting properties of neutron stars, 
and we will show how these conditions can be satisfied in our model. 
An important question which we wish to address is whether the resulting picture is consistent with
the generic stability conditions for a hybrid star, formulated in
Ref.~\cite{Alford:2013aca} and generalized in Ref.~\cite{Ranea-Sandoval:2015ldr}. 
Closely related to this is the role of the repulsive effects in quark matter
which arise from the interaction in the vector meson channels. 
The importance of the isoscalar-vector repulsion in producing a sufficiently stiff quark matter equation of state has been pointed out in several recent papers~\cite{Baym:2017whm,Whittenbury:2015ziz,Hell:2012da}, and here we will also include the isovector-vector repulsion which is very important in nuclei~\cite{Cloet:2009qs,Cloet:2012td}. We will compare the required strengths to those adjusted to the saturation density and symmetry energy of nuclear matter.

In order to gain insight into these and other related questions, we will 
confine ourselves to two light quark flavors in both the nuclear matter and the quark matter phases, 
and neglect finite size effects in the hadron-quark mixed phase. It is believed on rather general grounds that at very high
baryon densities a three-flavor color superconducting phase of quark matter is realized~\cite{Alford:2007xm}, 
but whether nuclear matter directly jumps to this phase or to an intermediate two-flavor color superconducting quark phase depends
on model details like the in-medium strange quark mass and the strength of the interaction in the relevant
diquark channels~\cite{Buballa:2003qv,Ruester:2005jc,Abuki:2005ms,Steiner:2002gx}. In any case, the role
of strangeness should be investigated consistently both in the hadronic phase, based for example
on the quark-diquark description of the baryon octet as given in Ref.~\cite{Carrillo-Serrano:2016igi}, 
as well as in the quark phase, and we wish to leave such extensions to a future work. 
Concerning finite size effects, it has been argued in several papers~\cite{Endo:2011em,Neumann:2002jm,Voskresensky:2002hu,Pinto:2012aq,Fraga:2018cvr} 
that surface tension and charge screening 
tend to work against spatially extended mixed 
phases, regaining qualitatively the simple picture of a Maxwellian first-order phase transition.
We will come back to this point in later sections. 

The outline of the paper is as follows: Sec.~\ref{sec:NJL} discusses the NJL model and the parameters that enter the calculations; Sec.~\ref{sec:matter} introduces the equations of state for nuclear matter and quark matter; Sec.~\ref{sec:results} presents a discussion of our results and a comparison with data and other related work; and a summary is given in Sec.~\ref{sec:summary}.

\section{LAGRANGIAN AND MODEL PARAMETERS\label{sec:NJL}}
The two-flavor NJL Lagrangian relevant for this study reads~\cite{Vogl:1991qt,Hatsuda:1994pi}:
\begin{align}
\mathcal{L} &= \widebar{\psi}\lf(i \slashed{\partial} - m\rg)\psi
+ G_{\pi} \left[ \big(\widebar{\psi} \psi\big)^{2} - \big(\widebar{\psi} \gamma_{5} \vec{\tau} \psi\big)^{2} \right]  \allowdisplaybreaks \nonumber \\
&- G_{\omega} \big(\widebar{\psi} \gamma^{\mu} \psi\big)^{2} 
 - G_{\rho} \big(\widebar{\psi} \gamma^{\mu} \vec{\tau} \psi\big)^{2} \nonumber \\ 
& + G_{S} \left(\widebar{\psi} \gamma_{5} C \tau_{2} \lambda^{A} \widebar{\psi}^{T}\right) 
          \left(\psi^{T} C^{- 1} \gamma_{5} \tau_{2} \lambda^{A} \psi\right),
\label{eq:lagrangian}
\end{align}
\looseness=-1
where $\psi$ is the quark field, $m$ is the current quark mass, $C$ is the charge conjugation matrix, $\lambda^A\,\,\,(A = 2, 5, 7)$ are the antisymmetric color Gell-Mann matrices, and $\vect{\tau}$ are the Pauli isospin matrices. The 4-fermion coupling constants in the scalar $\bar{q}q$ channel, the isoscalar and isovector vector $\bar{q}q$ channels, and the scalar $qq$ channel are denoted by $G_{\pi}$, $G_{\omega}$, $G_{\rho}$ and $G_S$, respectively.\footnote{The coupling constant $G_S$ of Eq.~\eqref{eq:lagrangian} is the same as $G_D$ used in most works on the NJL model for high density quark matter. It is related to $G_s$ used in our previous works on the NJL model (see for example Eq.~(7) of Ref.~\cite{Cloet:2009qs}) by
$G_S = \frac{3}{2} G_s$.}
The other model parameters are the 4-fermion coupling constant in the axial-vector $qq$ channel~\cite{Cloet:2014rja}, and the infrared (IR) and ultraviolet (UV) regularization parameters, 
which are used with the proper-time regularization scheme~\cite{Hellstern:1997nv,Bentz:2001vc}.\footnote{The interaction Lagrangian in the axial-vector diquark channel has the
form $G_{A} \left(\widebar{\psi} \gamma_{\mu} C \tau_i \tau_{2} \lambda^{A} \widebar{\psi}^{T}\right) \left(\psi^{T} C^{- 1} \gamma^{\mu} \tau_{2} \tau_i \lambda^{A}\psi\right)$, where $G_A$ is related to $G_a$ of our previous works (see for example Eq.~(7) of Ref.~\cite{Cloet:2009qs}) by $G_A = \frac{3}{2} G_a$. 
Because we do not consider spin triplet pairing here, 
this interaction is used only in the quark-diquark bound state equation.
We also mention that chiral symmetry of the interaction Lagrangian requires 
additional terms in the vector $qq$ and the 
axial-vector $\widebar{q}q$ channels, which however are not directly related to
our calculation.}
 
We stress that in this work we use the same model parameters as in several previous calculations which focused on the structure of nuclear matter systems~\cite{Cloet:2006bq,Cloet:2009qs,Cloet:2012td}. That is, the parameters of the model are determined in the vacuum, the single hadron sector, and nuclear matter sector as follows: We fix the IR cut-off $\Lambda_\mathrm{IR} = 0.24\,$GeV, and choose the UV cut-off $\Lambda_{\mathrm{UV}}$, $m$, and $G_\pi$ so as to give a constituent quark mass in vacuum of $M_0 = 0.4\,$GeV, the pion decay constant $f_\pi = 0.93\,$GeV, and the pion mass $m_\pi = 0.14\,$GeV using the standard methods based on the gap and Bethe-Salpeter equations~\cite{Cloet:2014rja}. The scalar diquark coupling $G_S$ and its axial-vector counterpart $G_A$ are then determined in the Faddeev equation approach to reproduce the vacuum values of the nucleon mass ($M_{N0} = 0.94\,$GeV) and the nucleon axial coupling ($g_{A0} = 1.267$)~\cite{Cloet:2005rt}. Finally, by using the model description for nuclear matter explained in Sec.~\ref{sec:matter}, the vector couplings $G_{\omega}$ and $G_{\rho}$ are determined from the binding energy per-nucleon in symmetric nuclear matter ($E_B = -15.7\,$MeV) and the symmetry energy ($a_4 = 32.0\,$MeV) at the saturation density~\cite{Cloet:2009qs}. We note that in this framework the saturation density of symmetric nuclear matter becomes $\rho_0 = 0.16\,$fm$^{-3}$. The resulting values for the model parameters are given in Tab.~\ref{tab:parameters}.\footnote{Note that $G_{\pi}$ obtained in the
present proper-time regularization scheme is almost three times larger than the
value obtained in the 3-momentum cut-off scheme (see for example Ref.~\cite{Buballa:2003qv}).
As a consequence, for the same vacuum value $M_0=0.4$ GeV, our vacuum chiral condensate
is roughly three times smaller in magnitude than in the
3-momentum cut-off scheme. Nevertheless, our value
$\langle \widebar{\psi} \psi \rangle_0^{1/3} = -0.216$ GeV is still close to the upper
limit of the range $[-0.33, -0.24]$ derived from QCD sum rules at a renormalization scale
of 1 GeV ~\cite{Dosch:1997wb}. By choosing smaller values for the input $M_0$ the magnitude of the chiral
condensate increases~\cite{Ninomiya:2014kja}.}
Using the nuclear matter equation of state presented in Sec.~\ref{sec:matter}, and the parameters of Tab.~\ref{tab:parameters}, gives an effective nucleon mass of $M_N = 0.744\,$GeV at nuclear matter saturation density, and an incompressibility of $K = 0.370\,$GeV.

In the quark matter phase there should be no effects from color confinement so we set 
$\Lambda_{\rm IR}=0$ in this phase. For the parameters $\Lambda_{\rm UV}$, $m$ and $G_{\pi}$,
which follow from the properties of the vacuum and the pion, we use the same values as in nuclear matter (see Tab.~\ref{tab:parameters}).  
As discussed in Sec.~\ref{sec:intro}, we wish to address the question whether the same
coupling constants $G_S$, $G_{\omega}$ and $G_{\rho}$ as determined by the free nucleon mass and
the properties of nuclear matter can also be used to describe the phase transition to color superconducting quark matter and neutron stars. We therefore introduce two scaling parameters 
$c_s$ and $c_v$ 
such that in the quark matter phase $G_S \to c_s\,G_S$, 
$G_{\omega} \rightarrow c_v \, G_{\omega}$, and $G_{\rho} \rightarrow c_v \, G_{\rho}$. 
(We use a common scaling factor for the vector-isoscalar and vector-isovector interaction in order to reduce the number of parameters.) The dependence of our results on the scaling parameters $(c_s, c_v)$ will be investigated
in detail in Sec.~\ref{sec:results}.

\section{NUCLEAR MATTER AND QUARK MATTER\label{sec:matter}}
In this section we present expressions for the effective potential ($V$) of nuclear matter (NM) and quark matter (QM) in the mean field approximation of the two-flavor  NJL model for given values of the two independent chemical potentials $\mu_B$ and $\mu_I$ for baryon number and isospin.
The corresponding chemical potentials for nucleons and quarks are\footnote{In principle a further chemical potential ($\mu_8$) is needed in QM to guarantee that the mean value of the color is zero, but it turns out to be very small for the two-flavor case~\cite{Buballa:2003qv,Aguilera:2004ag} and we neglect it here for simplicity.
We also note that in many works 
the chemical potentials of $u,d$ quarks are expressed as 
$\mu_q = \mu + Q_q \,\mu_Q$ ($q = u,d$), where $Q_q$ is the electric charge. 
Physical quantities as functions of baryon and
charge density of course do not change with different definitions of chemical potentials for baryon number
and charge.}
\begin{align}
\label{eq:mun}
\mu_p &= \mu_B + \mu_I,             & \mu_n &= \mu_B - \mu_I,   \\
\label{eq:muq}
\mu_u &= \frac{1}{3}\, \mu_B + \mu_I, & \mu_d &= \frac{1}{3}\, \mu_B - \mu_I.
\end{align}
The electron Fermi gas terms are also included, with the chemical potential $\mu_e = -2 \mu_I$ determined by $\beta$ equilibrium. Muon contributions will not be included for simplicity. 
Below we summarize the unregularized expressions, and refer to Ref.~\cite{Bentz:2001vc} for a detailed discussion on the proper-time regularization scheme.

The effective potential of NM for fixed chemical potentials $\mu_B, \mu_I$ is given by~\cite{Bentz:2002um,Lawley:2005ru}
\begin{align}
V^\mathrm{(NM)}(M, \omega_0, \rho_0) &= V_\mathrm{vac} + V_N - \frac{\omega_0^2}{4\,G_{\omega}} -
\frac{\rho_0^2}{4\,G_{\rho}} - \frac{\mu_e^4}{12 \pi^2},
\label{eq:V_NM}
\end{align}
where $M$ is the (in-medium) constituent quark mass and  
$\omega_0$ and $\rho_0$ are the isoscalar- and isovector-vector mean fields,
which must be determined by minimization of the effective potential.
Those minimization conditions are equivalent to the following definitions in terms of the in-medium 
quark condensates:
\begin{align}
M &= m - 2\,G_{\pi} \langle \widebar{\psi} \psi \rangle \,, 
\label{eq:gap} \\
\omega_0 &= 2\, G_{\omega} \langle \psi^{\dagger} \psi \rangle \,,
\,\,\,\,\,\,\,\,\,\,\,\,\rho_0 = 2\, G_{\rho} \langle \psi^{\dagger} \tau_3 \psi \rangle \,. 
\label{eq:vec}
\end{align}
The vacuum (Mexican hat shaped) contribution is
\begin{align}
V_\mathrm{vac} = 12 \, i \int \frac{\mathrm{d}^4 k}{(2 \pi)^4}\, \ln \frac{k^2 - M^2}{k^2 - M_0^2}
+ \frac{(M-m)^2}{4\,G_{\pi}} - \frac{(M_0-m)^2}{4\,G_{\pi}}.
\label{eq:V_vac}
\end{align}
The Fermi motion of nucleons in the scalar and vector
mean fields is described by the term
\begin{align}
\hs*{-2mm}V_N = -2 \sum_{\alpha = p,n} \int \frac{{\rm d}^3 k}{(2\pi)^3}
    \left(\mu_{\alpha}^* - E_N(k)\right) \
    \Theta\left(\mu_{\alpha}^* - E_N(k)\right),
    \label{eq:V_N}
\end{align}
where the effective chemical potential for protons and neutrons is given by
\begin{align}
    \mu_{\alpha}^{*} = \mu_{\alpha} - 3 \omega_0 \mp \rho_0 \,\,\,\,\,\, (\alpha = p, n),
    \label{eq:eff_mu_alpha}
\end{align}
and $E_{N}(k) = \sqrt{\vect{k}^{2} + M_{N}^{2}}$. The nucleon mass $M_N(M)$ is determined as a function of the constituent quark mass $M$ from the relativistic Faddeev equation for the nucleon, which is approximated as a quark-diquark bound state~\cite{Cloet:2014rja}. As mentioned in Sec.~\ref{sec:intro}, the function $M_N(M)$ develops a (positive) curvature with decreasing $M$, which is crucial for the saturation of the NM binding energy~\cite{Bentz:2001vc}.

\begin{table}[tbp]
\addtolength{\extrarowheight}{2.2pt}
\centering
\caption{Values for the model parameters which are determined in the vacuum, single hadron, and nuclear matter sectors. The regularization parameters and the current quark mass are given in units of GeV, and the coupling constants in units of GeV$^{-2}$.}
\begin{tabular*}{\columnwidth}{@{\extracolsep{\fill}}cccccccc}
\hline\hline
$\Lambda_\mathrm{IR}$  & $\Lambda_{\rm UV}$  &  $G_{\pi}$  &  $G_S$  &  $G_A$  &  $G_{\omega}$ &  $G_{\rho}$ & $m$ \\
\hline
0.240 & 0.645 & 19.04 & 11.24 & 4.20 & 6.03 & 14.17 & 0.016  \\
\hline\hline
\end{tabular*}
\label{tab:parameters}
\end{table}

The effective potential for QM for fixed chemical potentials $\mu_B$ and $\mu_I$ is given by~\cite{Bentz:2002um,Lawley:2005ru}
\begin{align}
    V^\mathrm{(QM)}(M,\Delta, \omega_0, \rho_0) = V_\mathrm{vac} + V_{Q} + V_{\D}
        - \frac{\omega_0^2}{4\,G_{\omega}}
        - \frac{\rho_0^2}{4\,G_{\rho}}
        - \frac{\mu_e^4}{12 \pi^2},
    \label{eq:V_QM}
\end{align}
where the constituent quark mass $M$ and the mean vector fields $\omega_0$ and $\rho_0$ in QM 
are given in terms of quark condensates as in Eqs.~\eqref{eq:gap} and~\eqref{eq:vec},
while the energy gap $\Delta$ arising from the pairing in the scalar diquark channel is related to the BCS-type quark condensate (order parameter of the broken color symmetry) by
\begin{align}
\Delta = - 2\,G_S\,\langle \psi^{T} \, C \gamma_5 \tau_2 \lambda_2 \psi \rangle \,.
\label{eq:bcs}
\end{align}
It is well known~\cite{Shovkovy:2004me,Alford:2007xm} that this type of 2-flavor pairing leaves chiral symmetry intact, and
in the limit $m=0$ color superconducting quark matter may therefore exist in the chiral
symmetric phase ($M=0$). 
All four quantities $M, \Delta, \omega_0, \rho_0$ are determined by minimization of the effective potential for fixed chemical potentials. 

The vacuum part $V_\mathrm{vac}$ in Eq.~\eqref{eq:V_QM} is the same as given in Eq.~\eqref{eq:V_vac}. The difference between its value at $M$ calculated in QM and
its value at $M$ calculated in NM corresponds to the bag constant in the NJL model.  
The term $V_{Q}$ describes the Fermi motion of quarks moving in the scalar and vector mean fields:
\begin{align}
    V_\mathrm{Q} = -6 \sum_{\alpha = u,d} \int \frac{{\rm d}^3 k}{(2\pi)^3}
         \left(\mu_{\alpha}^* - E(k)\right)
        \ \Theta\left(\mu_{\alpha}^* - E(k)\right).
    \label{eq:V_Q}
\end{align}
Here $E(k) = \sqrt{\vect{k}^2 + M^2}$ and the effective up and down quark chemical
potentials are defined by
\begin{align}
    \mu_{\alpha}^{*} = \mu_{\alpha} - \omega_0 \mp \rho_0 \,\,\,\,\,\, (\alpha = u, d).
    \label{eq:eff_mu_alpha_QM}
\end{align}
The term $V_{\Delta}$ in Eq.~\eqref{eq:V_QM} arises from the pairing in the scalar diquark channel, and is given by (see the papers on color superconductivity cited in Sec.~\ref{sec:intro} and also Refs.~\cite{Kiriyama:2001ud,Huang:2003xd})
\begin{align}
V_{\Delta} &= 2 i \int \frac{{\rm d}^4 k}{(2 \pi)^4}
        \sum_{\alpha=+,-}
        \left[ \ln \frac{ k_0^2 - \left(\epsilon_{\alpha} + \mu_I^* \right)^2}
            { k_0^2 - \left(E_{\alpha} + \mu_I^* \right)^2} \right. \nonumber \\
&\hs*{27mm}
\left. + \ln \frac{ k_0^2 - \left(\epsilon_{\alpha} - \mu_I^* \right)^2}
            { k_0^2 - \left(E_{\alpha} - \mu_I^* \right)^2}
        \right] + \frac{\Delta^2}{4\,G_S},
    \label{eq:V_Delta}
\end{align}
where $\epsilon_{\pm} = \sqrt{ \left[E(k) \pm \mu_B^*/3 \right]^2 + \Delta^2}$ is the quark dispersion relation\footnote{As can be seen from our results in Sec.~\ref{sec:results}, we always have
$\Delta > |\mu_I^*|$, i.e., the ``gapless modes''~\cite{Shovkovy:2004me}  
(poles at zero frequency for finite $\Delta$) are not realized in our calculation, which is consistent with the findings of other works~\cite{Ruester:2005jc} for intermediate or strong pairing strength.}
and $E_{\pm} = \left|E(k) \pm \mu_B^*/3\right|$. 
The effective chemical potentials for baryon number and isospin are
\begin{align}
\mu_B^* &= \mu_B - 3\,\omega_0,  &  \mu_I^*&= \mu_I -  \rho_0.
\end{align}
Using the above forms for the effective potentials, the pressure ($P$), baryon and isospin densities ($\rho_B$, $\rho_I$), and the energy density (${\cal E}$) are obtained by
\begin{align}
\label{diff} 
P &= - V, & \rho_{\alpha} &= -\frac{\partial V}{\partial \mu_{\alpha}} \,\,\,\,\,(\alpha=B, I),  \\
{\cal E} &= V + \sum_{\alpha=B,I} \mu_{\alpha} \rho_{\alpha}.
\end{align}
The charge neutrality condition $\rho_C = \left(\rho_B + \rho_I\right)/2 =0$ then implies a relation
between the two chemical potentials $\mu_B$ and $\mu_I$. The charge neutral equation
of state is then a function of only one variable, which we take to be the baryon density $\rho_B$.

The equation of state for the globally charge neutral mixed phase is then 
calculated by using the Gibbs
construction~\cite{Glendenning:1992vb} as follows:
If there is a line in the $(\mu_B, \mu_I)$ plane between two points (called X and Y) along which
the NM and QM phases have equal effective potentials (denoted as $V^{\rm (mixed)}$) and opposite charges, then along this line we
have trivially
$V^{\rm (mixed)} = x_1 \, V^{\rm (NM)} + (1-x_1) \, V^{\rm (QM)}$ for any number $x_1$. We can calculate the
densities and the energy density along this line by differentiation of $V^{\rm (mixed)}$ according to Eq.~\eqref{diff},
noting that
$\partial V^{\rm (mixed)}/\partial x_1=0$. The requirement that the charge density along this line vanishes, i.e.,
$\rho_C = x_1 \rho_C^{\rm (NM)} + (1-x_1) \rho_C^{\rm (QM)}=0$, determines $x_1(\mu_B, \mu_I)$ as the
volume fraction of NM in the mixed phase as
\begin{align}
x_1(\mu_B, \mu_I) &= \frac{\rho_C^{\rm (QM)}}{\rho_C^{\rm (QM)} - \rho_C^{\rm (NM)}}.
\end{align}
If we approach the point X along a charge neutral line of the NM phase,
then $x_1 =1$ at point X, and if we leave point Y towards a charge neutral line of the QM phase,
then $x_1 = 0$ at point Y, i.e., $x_1$ decreases from $1$ to $0$ along the line X~$\rightarrow$~Y.
The baryon density in the mixed phase is then given
by $\rho_B^{\rm (mixed)} = x_1 \rho_B^{\rm (NM)} + (1-x_1) \rho_B^{\rm (QM)}$, and a similar expression holds also
for the energy density.

\section{RESULTS AND DISCUSSION\label{sec:results}}
Before presenting our results, we would like to explain how they qualitatively depend on the choice of
parameters.   
The equation of state in the QM phase is largely controlled by three parameters in our model: the strength of the attractive isoscalar-scalar pairing interaction ($G_S$) and the strengths of the repulsive isoscalar- and isovector-vector interactions $G_{\omega}$, $G_{\rho}$. If we increase the pairing attraction, the baryon density $\rho_{\rm tr}$ where the transition NM $\rightarrow$ QM occurs decreases, resulting in an overall softening of the equation of state and lower masses of neutron stars. If the pairing interaction is increased beyond a certain limit, QM becomes the stable phase of the system also at low densities, which we regard as unphysical. On the other hand, if we increase the vector repulsion the transition density $\rho_{\rm tr}$ increases, then the equation of state becomes stiffer and the masses of the neutron stars increase. If the vector repulsion is increased beyond a certain limit, NM becomes the stable phase also for high densities, which we again regard as unphysical.

As explained at the end of Sec.~\ref{sec:NJL}, in order to show these dependencies quantitatively we 
scale the value of $G_S$ given in Tab.~\ref{tab:parameters} by a factor $c_s$, and the values of
$G_{\omega}$ and $G_{\rho}$ by a common factor $c_v$, in the calculation of the QM sector.  
To characterize the results obtained for different values of $c_s$ and $c_v$ we define a ``physically reasonable'' scenario by the following three conditions: (1) The phase transition NM $\rightarrow$ QM occurs in the range $2\,\rho_0 \leq \rho_{\rm tr} \leq 4\,\rho_0$;\footnote{There is no fundamental reason for this choice of limits and our qualitative results do not change much if this condition is relaxed.} (2) the maximum mass of the star satisfies $M_{\rm star}^{\rm (max)} \geq  1.97\,M_\odot$ to be compatible with recent observations~\cite{Demorest:2010bx,Antoniadis:2013pzd}; and (3) the hybrid star with a 
superconducting QM core is stable against density fluctuations, i.e., ${\rm d}M_{\rm star}/{\rm d} \rho_B(r=0) > 0$ where $\rho_B(r=0)$ is the central baryon density, in a region of densities above the transition density. 

\begin{figure}[tbp]
\centering\includegraphics[clip, width=\columnwidth]{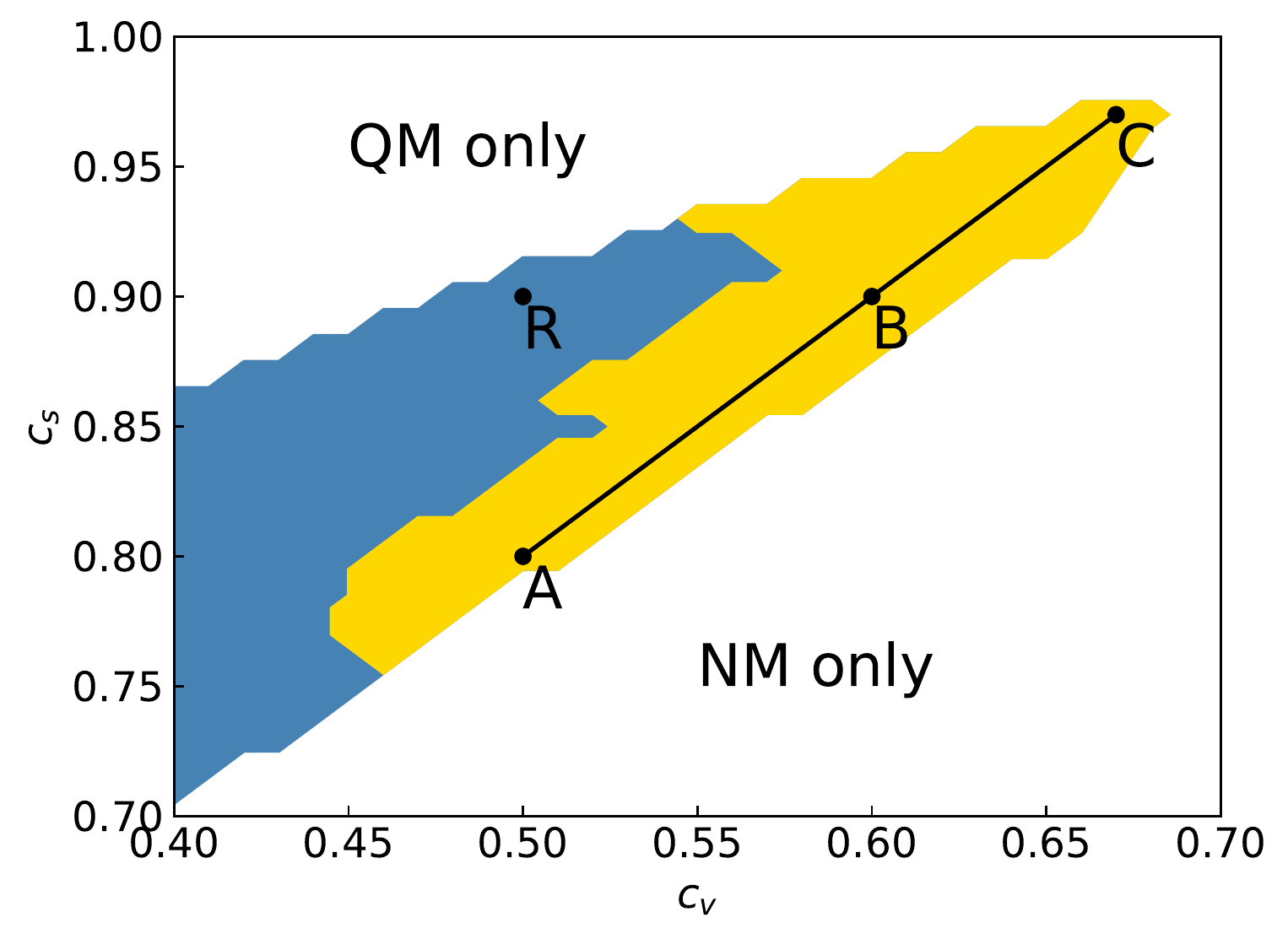}
\caption{(Color online) Scaling factors $c_s$ and $c_v$ for the interactions in the scalar diquark
    and vector meson channels, used in the QM sector of the calculation.
    The conditions (1), (2), and (3) specified in the text are satisfied in the
    yellow region, which contains the straight line $c_s = c_v + 0.3$ connecting the points A, B and C. 
    In the blue region, containing the reference points R, at least one of the conditions (2) or (3) is not satisfied. 
    In the white region marked as ``NM only'', either the transition density is too high or only the NM phase is 
    realized, while in the white region marked as ``QM only'', either the transition density is too low or
    only the QM phase is realized.}
    \label{fig:map}
\end{figure}

We find that all three conditions can be satisfied in the rather narrow parameter region marked by yellow in 
Fig.~\ref{fig:map}. In the blue region of Fig.~\ref{fig:map}, which continues towards smaller values of $(c_s, c_v)$ not included in the
figure, at least one of the conditions (2) or (3) is not satisfied.   
Because the yellow region extends up to $c_s=0.98$, we find that in practice it is possible to use the same value of $G_S$ for the scalar pairing strength in QM as for the scalar diquark interaction in the single nucleon sector. On the other hand, if we would use the same value of the vector couplings $G_{\omega}$, $G_{\rho}$ as determined in the nuclear matter sector (see Tab.~\ref{tab:parameters}), there would be no phase transition to QM. In order to satisfy the three conditions explained above, the vector coupling in QM must be smaller than in NM by a factor of $0.45 \leq c_v \leq 0.68$.\footnote{Interestingly, using the same NJL framework to explore possible explanations of the EMC effect~\cite{Aubert:1983xm,Geesaman:1995yd,Malace:2014uea,Cloet:2019mql} in nuclear structure functions, we found that the coupling of the vector mean field to the struck quark must also be substantially reduced~\cite{Cloet:2006bq}. In this case the reduction was associated with asymptotic freedom in QCD~\cite{Detmold:2005cb}.}
In the white region marked as ``NM only'', either the transition
density is too high or only the NM phase is realized, while in the
white region marked as ``QM only'', either the transition density is too low or
only the QM phase is realized.

\begin{figure}[tbp]
  \centering\includegraphics[width=\columnwidth]{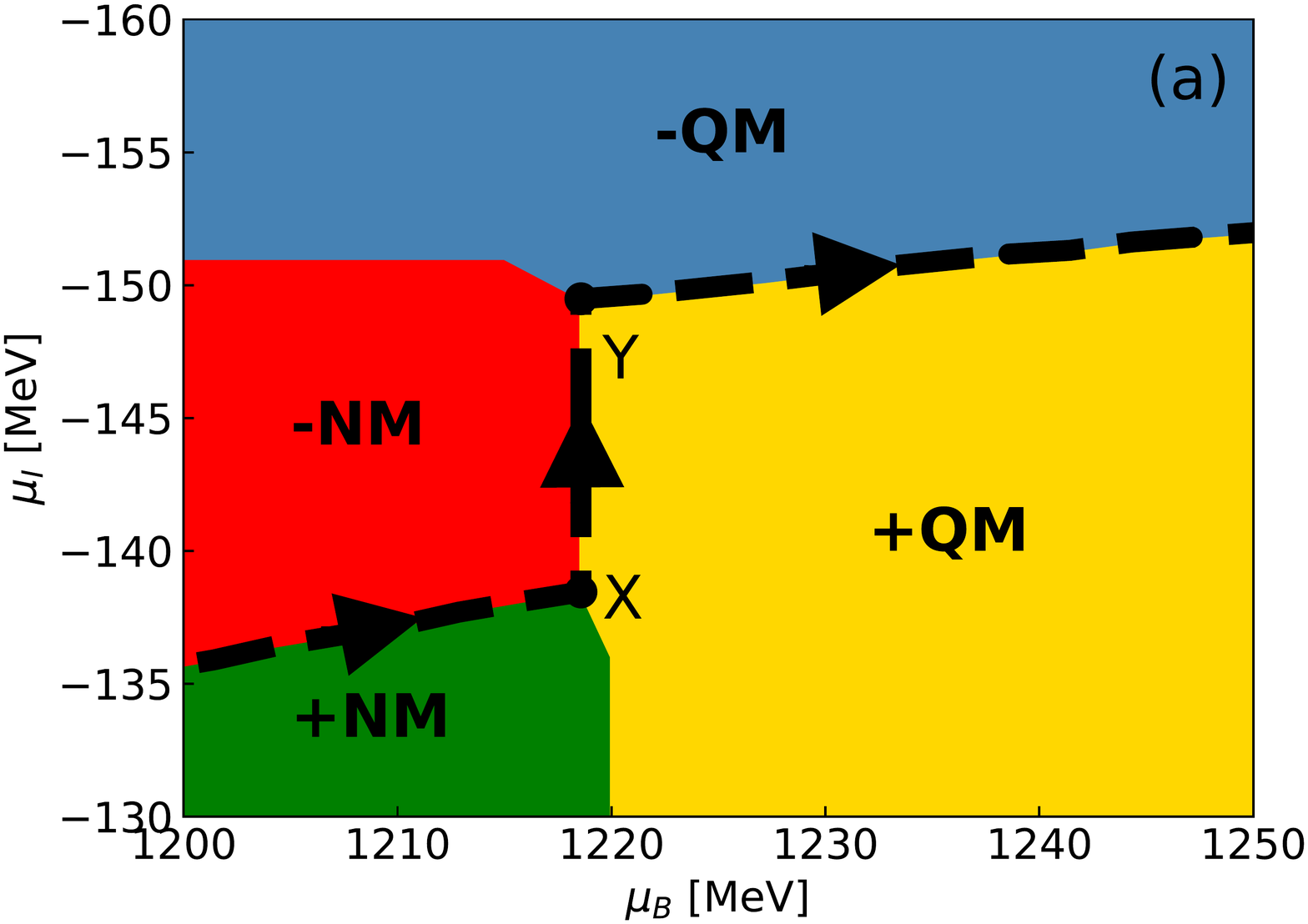} \\[1.0em]
  \centering\includegraphics[width=\columnwidth]{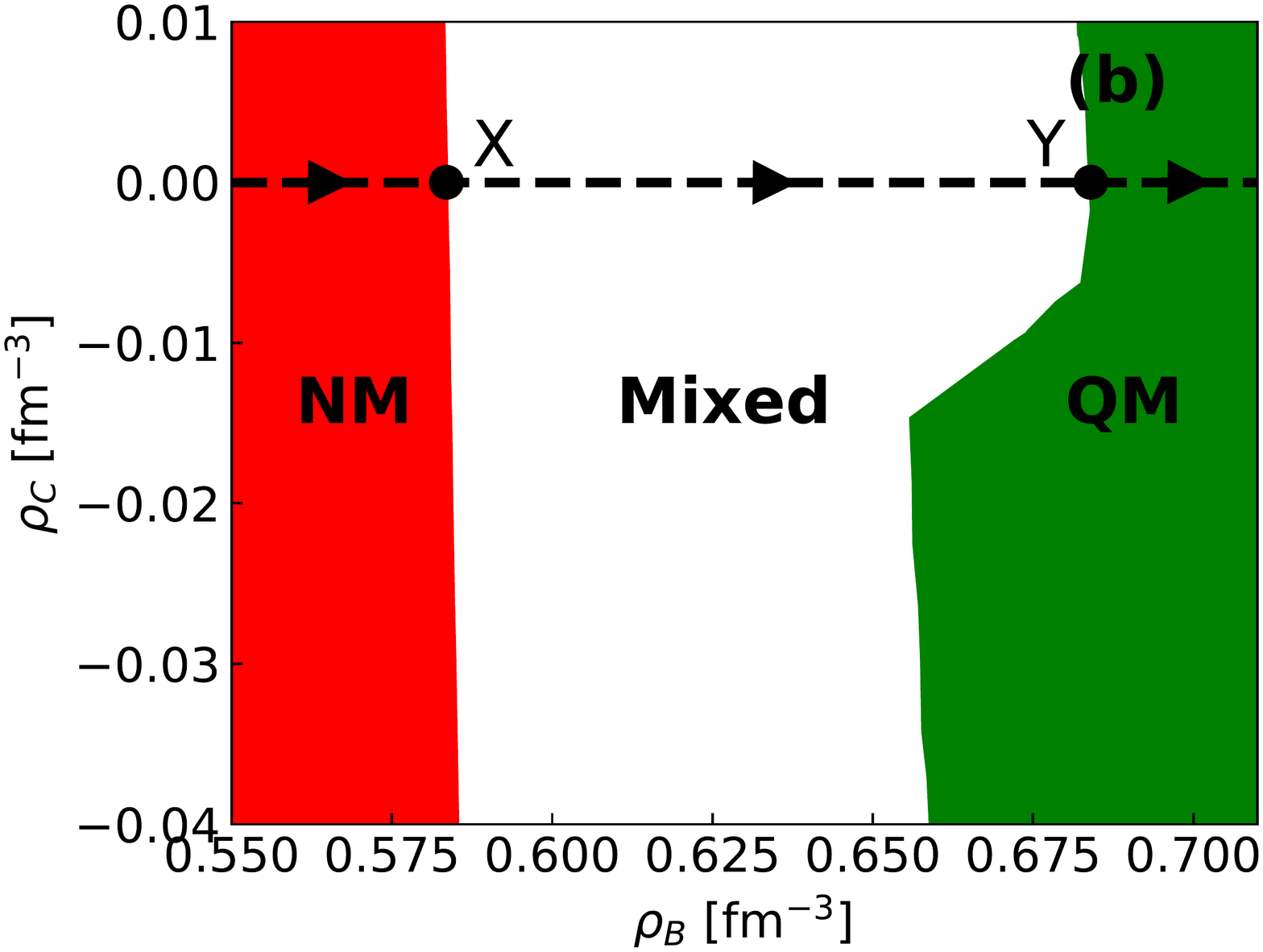}
\caption{(Color online) Phase diagrams for the point B of Fig.~\ref{fig:map} ($c_s = 0.9, c_v = 0.6$) in: (a) the plane of chemical potentials, and (b) in the plane of densities. In the top panel (a) the 
``$+$'' and ``$-$'' signs refer to the electric charge state of nuclear matter (NM) or quark matter (QM). 
The black dashed line markes electrically neutral matter, with the arrows indicating the increase
of baryon density. The segment X~$\rightarrow$~Y markes the mixed phase.}
\label{fig:phase_diagram}
\end{figure}

In order to discuss our results for the phase structure, we select point B of 
Fig.~\ref{fig:map} as a representative example where all three conditions are satisfied.
Figs.~\ref{fig:phase_diagram} shows the phase diagrams in: (a) the plane of chemical potentials, and (b) the plane of densities,
focusing on the region of the phase transition for electrically neutral matter.
In Fig.~\ref{fig:phase_diagram}\textcolor{blue}{a} the stable phase (phase with the larger pressure) is marked as NM or QM for each point in the $(\mu_B, \mu_I)$ plane, together with the sign of the electric charge density. The dashed line marks electrically neutral matter. The baryon density increases as we follow this line from the left end in the NM phase to the right end in the QM phase. We see that in the section of the NM-QM mixed phase\emdash
the line X~$\rightarrow$~Y in Fig.~\ref{fig:phase_diagram}\textcolor{blue}{a}\emdash $\mu_B$ stays almost constant
(within our numerical accuracy of $\pm 1$ MeV),  
while $|\mu_I\!|$ increases by about 10 MeV.\footnote{Note, however, that the constancy 
(almost zero width) of $\mu_B$ during the phase transition depends somewhat on the 
definitions of the chemical potentials, see Footnote 4. With our present definitions,
we find very small widths ($< 3$ MeV) for most of the parameters $(c_s, c_v)$ in the colored regions
of Fig.~\ref{fig:map}.
Large widths occur only for weak pairing strength and weak vector interaction ($c_s \lesssim 0.2$, $c_v \simeq 0$),
which is consistent with the results of Refs.~\cite{Bentz:2002um,Lawley:2005ru}, but in this region
the conditions (2) and (3) cannot be satisfied at the same time.}

Fig.~\ref{fig:phase_diagram}\textcolor{blue}{b} shows the phase diagram in the plane of the densities 
$(\rho_B, \rho_C)$
in the vicinity of $\rho_C = 0$. The dashed line, which marks electrically neutral matter, passes through the region of the mixed phase (shown in white). Figs.~\ref{fig:phase_diagram} illustrates that along the phase transition line X~$\rightarrow$~Y there is only
small change of the chemical potentials, while the change in baryon density (about $0.1$ fm$^{-3}$) is appreciable but not too large, so as to keep our resulting hybrid star gravitationally stable as will be shown below. 
We found qualitatively very similar results for the phase diagrams for all parameter sets $(c_s, c_v)$ in the colored regions of Fig.~\ref{fig:map}.  

\begin{figure}[tbp]
  \centering\includegraphics[width=\columnwidth]{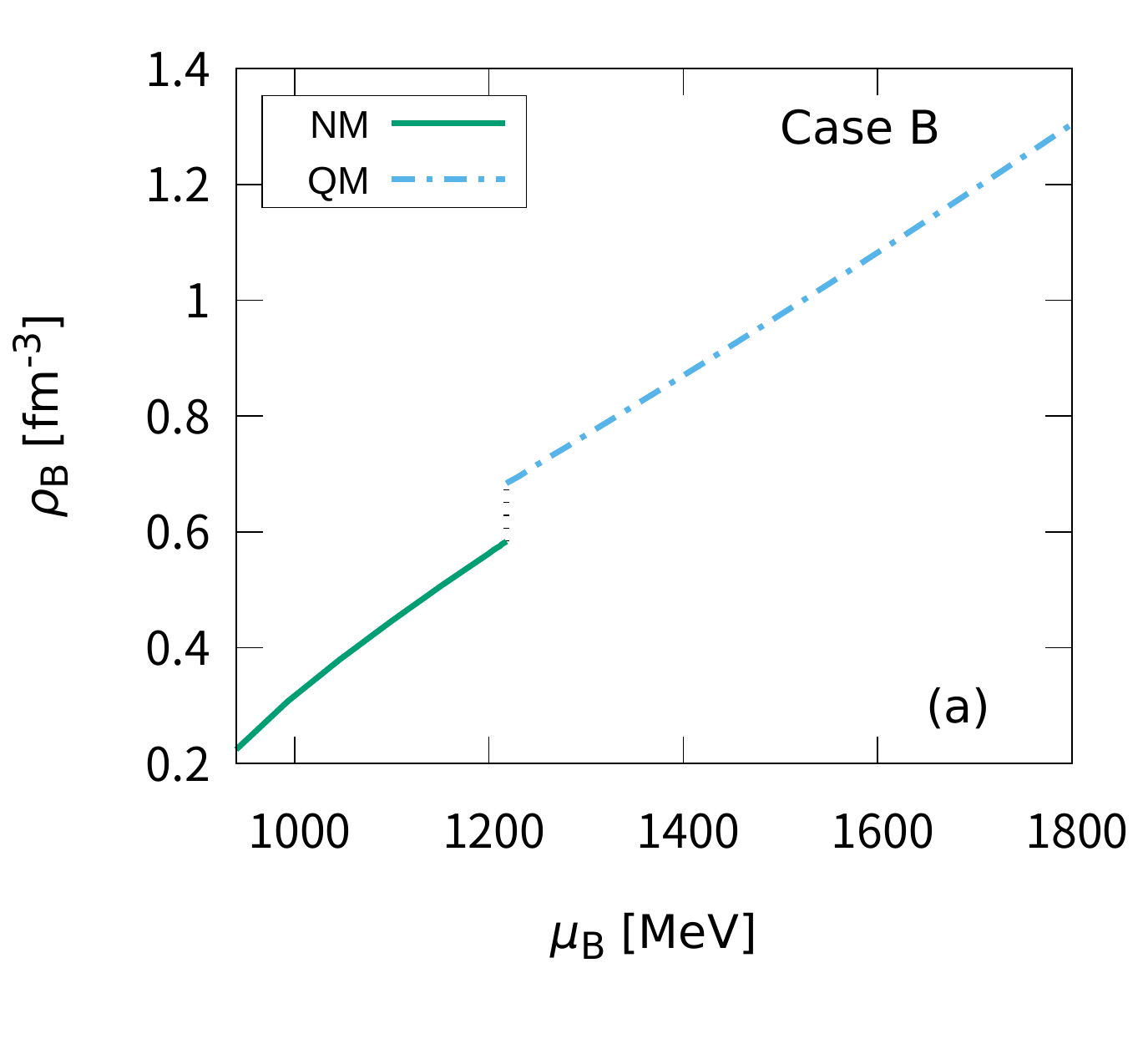} \\
  \centering\includegraphics[width=\columnwidth]{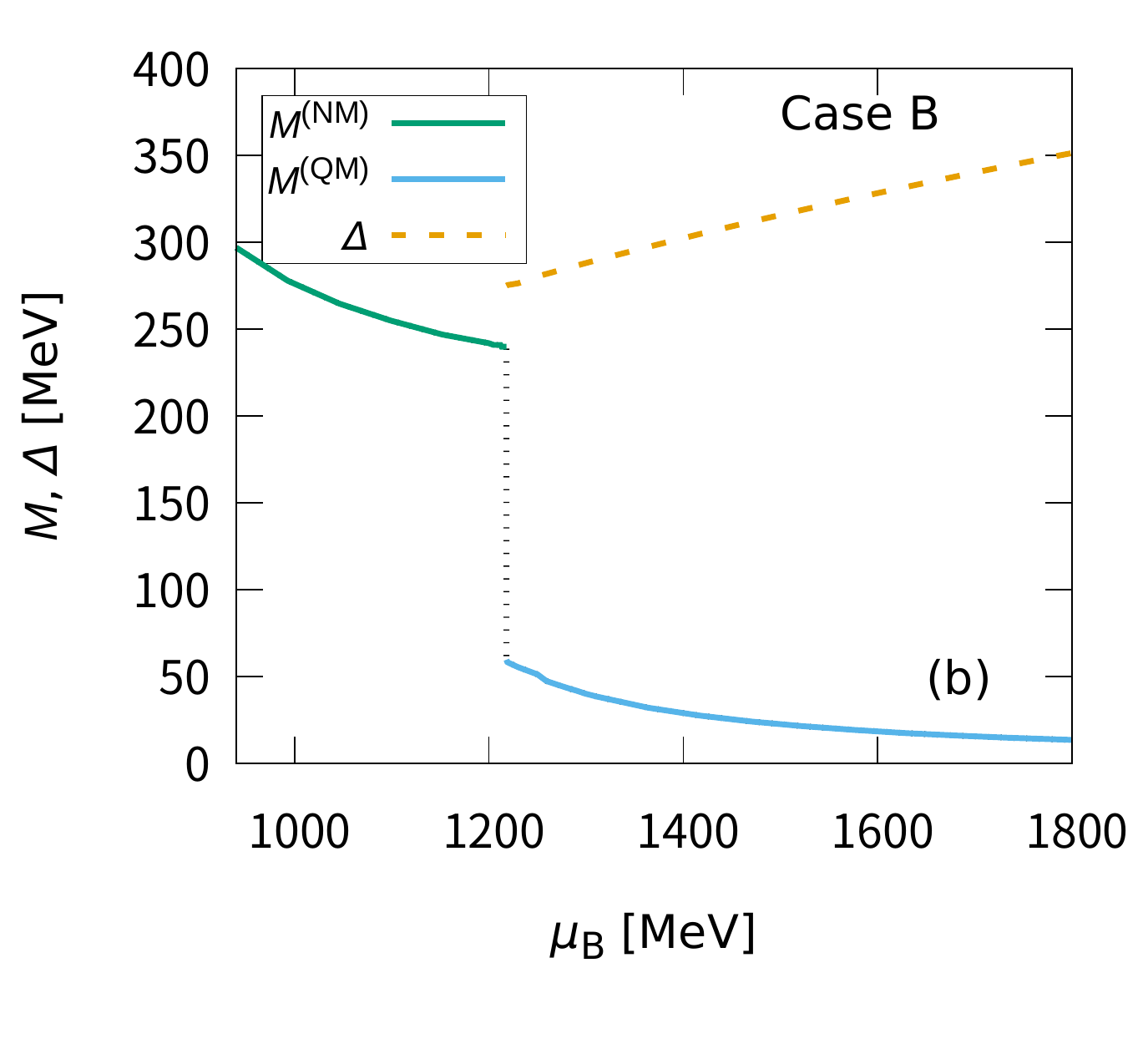}
\caption{(Color online) The dependence of physical quantities on the baryon chemical potential in electrically neutral matter for the case of point B in Fig.~\ref{fig:map} ($c_s = 0.9, c_v = 0.6$). Figure (a) shows the baryon density and (b) shows the constituent quark mass $M$ and the energy gap $\Delta$.}
\label{fig:mu_B-physval}
\end{figure}

In Fig.~\ref{fig:mu_B-physval}\textcolor{blue}{a} we show the baryon density of electrically neutral matter as function of the baryon chemical potential for the parameter set B. As can be anticipated from Figs.~\ref{fig:phase_diagram}, the change of the baryon density during the phase transition appears as a jump, 
i.e., within our numerical accuracy (of $\pm 1$ MeV) it occurs at constant
$\mu_B$. In Fig.~\ref{fig:mu_B-physval}\textcolor{blue}{b} we show the constituent quark mass $M$ and the energy gap $\Delta$ as functions of $\mu_B$ for the same parameter set B. The values of $M$ in the QM phase are small compared to those in the
vacuum or the NM phase, which indicates that QM in the range of $\mu_B$ above the phase transition is already reasonably close to a phase
where chiral symmetry is restored. Our values of $\Delta$ are larger than in most of the previous works
done in the 3-momentum cut-off scheme (see for example Refs.~\cite{Buballa:2003qv,Kiriyama:2001ud}), which is mainly because of the larger coupling constant 
$G_S$ in Eq.~\eqref{eq:bcs} for the proper-time regularization scheme, see Tab.~\ref{tab:parameters} for the reference value. (Our values for the BCS-type condensate in Eq.~\eqref{eq:bcs} are very similar to the values obtained with the 3-momentum cut-off scheme.) Nevertheless, in the region of $\mu_B$ just above the
phase transition, we observe a qualitative agreement of our results shown in Fig.~\ref{fig:mu_B-physval}\textcolor{blue}{b} with the results for the two-flavor quark phase shown in Fig.~33 of Ref.~\cite{Baym:2017whm}, or Fig.~6.8 of Ref.~\cite{Buballa:2003qv}. We found that all parameter sets $(c_s, c_v)$ shown by the yellow region in Fig.~\ref{fig:map} give results which are qualitatively very similar to those shown in Figs.~\ref{fig:mu_B-physval}.         

\begin{figure}[tbp]
  \centering\includegraphics[width=\columnwidth]{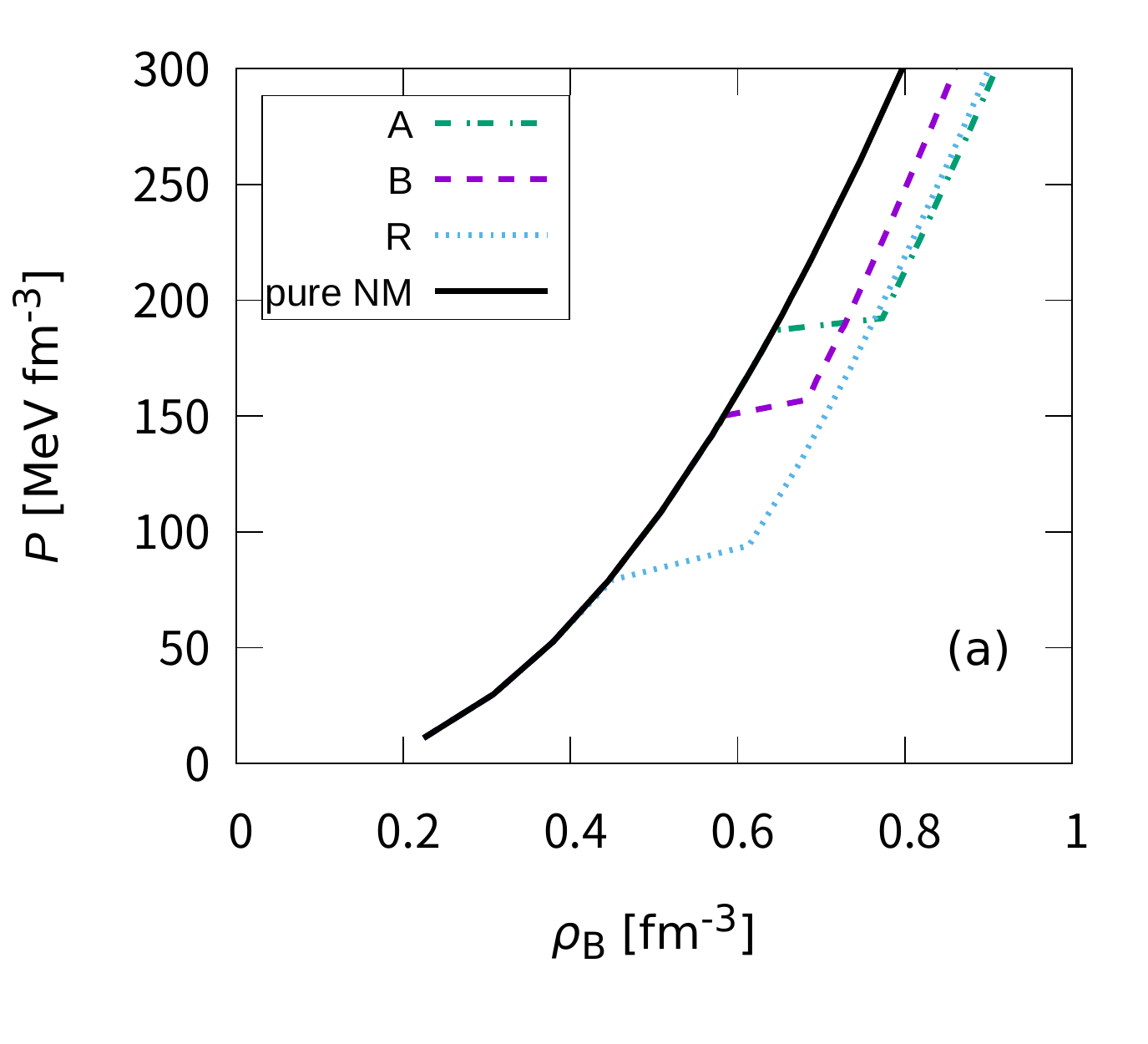} \\
  \centering\includegraphics[width=\columnwidth]{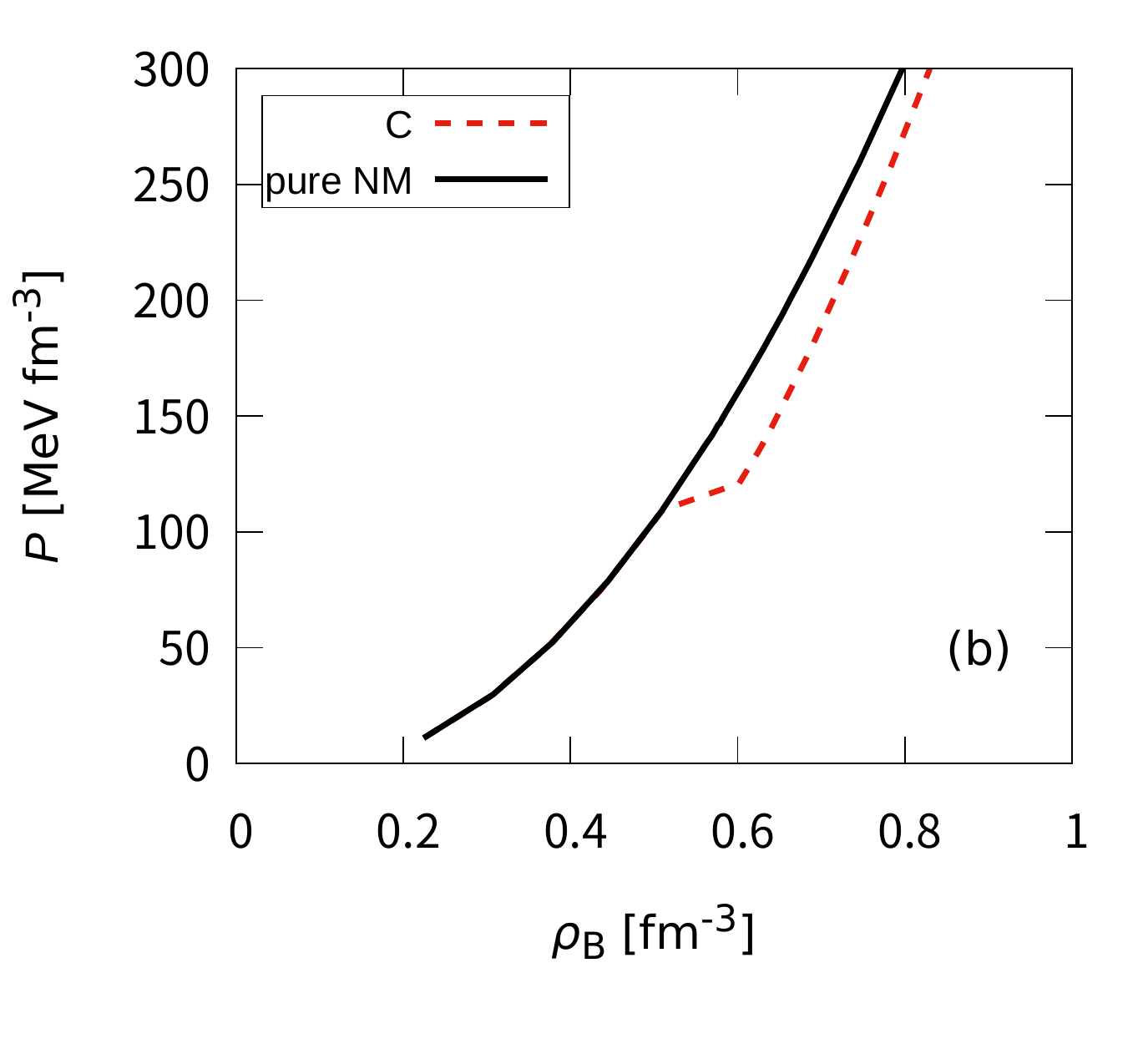}
\caption{(Color online) Pressure as function of the baryon density for electrically neutral matter. Figure (a) shows the results obtained for the points A and B in comparison to point R of Fig.~\ref{fig:map}, and (b) 
shows the result obtained for the point C. The black solid line is the result of the pure NM case.}
\label{fig:p-rho_B}
\end{figure}

In order to separately show the counteracting effects of pairing and vector repulsion in the QM phase, 
we show in Fig.~\ref{fig:p-rho_B}\textcolor{blue}{a} the pressure of electrically neutral matter as a function
of the baryon density for the points A, B and R in Fig.~\ref{fig:map}, as well as the result for the pure NM case. Here the point R, which satisfies the
above conditions (1) and (3) but not (2), is used as a reference point.   
Starting with case A, we see that by increasing the pairing strength (${\rm A} \rightarrow {\rm R}$) the transition density decreases substantially, resulting in an overall softer equation of state. If the vector coupling is then increased (${\rm R} \rightarrow {\rm B}$) the transition density increases again, but stays below case A. As a result, there is a net decrease of the transition density and a resulting softening of the equation of state as we go along the solid line of Fig.~\ref{fig:map} in the direction of increasing coupling constants ({\rm A} $\rightarrow$ {\rm B}). 
This trend continues as we extend the line $c_s = c_v + 0.3$ up to the point C, for which the results are shown in Fig.~\ref{fig:p-rho_B}\textcolor{blue}{b}. In this case $c_s = 0.98$, i.e., the coupling constant $G_S$ is practically the same as determined from the free nucleon mass and the vector couplings $G_{\omega}$, $G_{\rho}$ are smaller than the nuclear matter values by a factor of $c_v = 0.68$. The transition density is about $3\,\rho_0$ for this case.

The results shown in Figs.~\ref{fig:p-rho_B} indicate that the phase transition, which we described here by the Gibbs construction, is in fact very similar to the usual Maxwellian first order phase transition. This result, which for the case B can be anticipated from Figs.~\ref{fig:phase_diagram} and~\ref{fig:mu_B-physval}, 
holds for the whole yellow region of parameters shown in Fig.~\ref{fig:map}. As a consequence, 
the spatial extension of the mixed phase in a hybrid star will be small compared to the
overall size, i.e., if the central density is sufficiently high the QM phase will begin to form at
the center of the star with an almost sharp boundary to the surrounding NM phase, which is the scenario depicted schematically in Fig.~25 of Ref.~\cite{Heiselberg:1999mq}. Our results for the phase transition also suggest that the inclusion of finite size effects (surface tension of various
geometrical shapes, Coulomb interactions and charge screening) in the
mixed phase will not lead to qualitative changes of the overall physical picture: As mentioned in Sec.~\ref{sec:intro}, these effects work against spatially extended mixed 
phases, regaining qualitatively the simple picture of a Maxwellian first-order phase transition
~\cite{Endo:2011em,Neumann:2002jm,Voskresensky:2002hu}.

\begin{figure}[tbp]
\centering\includegraphics[clip, width=\columnwidth]{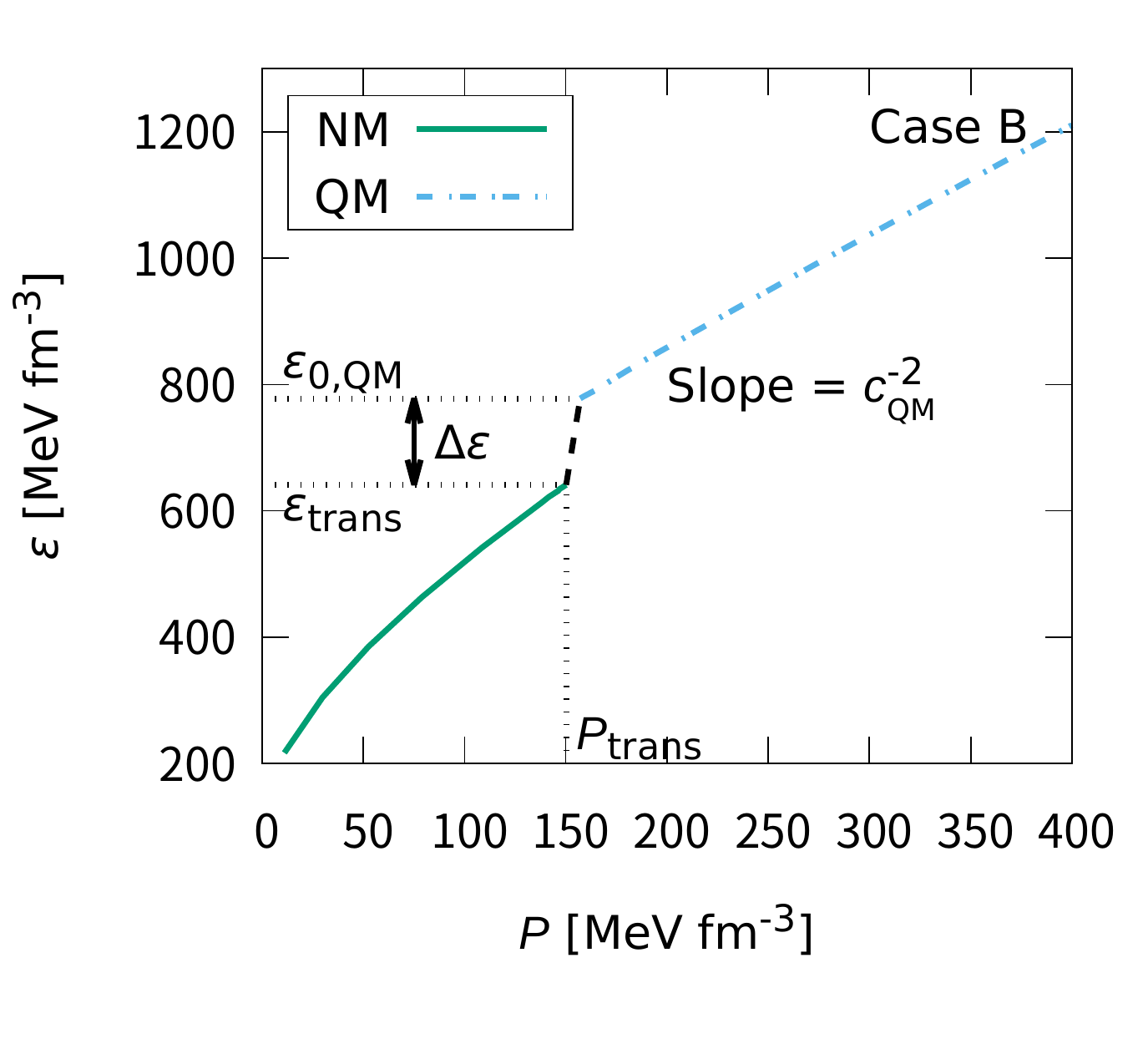}
\caption{(Color online) Equation of state (energy density vs. pressure) of charge neutral matter for the case 
of parameter set B of Fig.~\ref{fig:map}. $P_{\rm trans}$ and ${\cal E}_{\rm trans}$ are the pressure and
energy density of the transition in the NM phase, ${\cal E}_{0, \rm{QM}}$ is the energy density
at the beginning of the QM phase, and $\Delta {\cal E} = {\cal E}_{0, \rm{QM}} - {\cal E}_{\rm trans}$. 
The squared speed of sound in the QM phase is 
$c_{\rm QM}^2 = \left({\rm d}P / {\rm d} {\cal E}\right)_{\rm QM}$.} 
\label{fig:alford}
\end{figure}

In order to see whether our results are consistent with phenomenological parametrizations based
on generic stability criteria for hybrid stars~\cite{Alford:2013aca,Ranea-Sandoval:2015ldr}, we show
our equation of state (energy density vs. pressure) for case B in Fig.~\ref{fig:alford}.
We immediately see that the physical picture is consistent with those parametrizations:
The increase of the energy density during the phase transition ($\Delta {\cal E} = 136.7$ MeV/fm$^3$) is 
smaller than the critical value determined from Eq.~(2) of Ref.~\cite{Alford:2013aca}
($\Delta {\cal E}_{\rm crit}  = 545.9$ MeV/fm$^3$ in our case), 
which means that the pressure of QM can counteract the additional downward gravitational pull\emdash
exerted by the additional energy in the core\emdash on the rest of the star. 
We can see also consistency with the more general parametrizations based on a constant speed of sound in QM~\cite{Ranea-Sandoval:2015ldr}, and our value $c_{\rm QM}^2 = 0.56$ is sufficiently high
to get a heavy hybrid star~\cite{Alford:2013aca}.

\begin{figure}[tbp]
  \centering\includegraphics[width=\columnwidth]{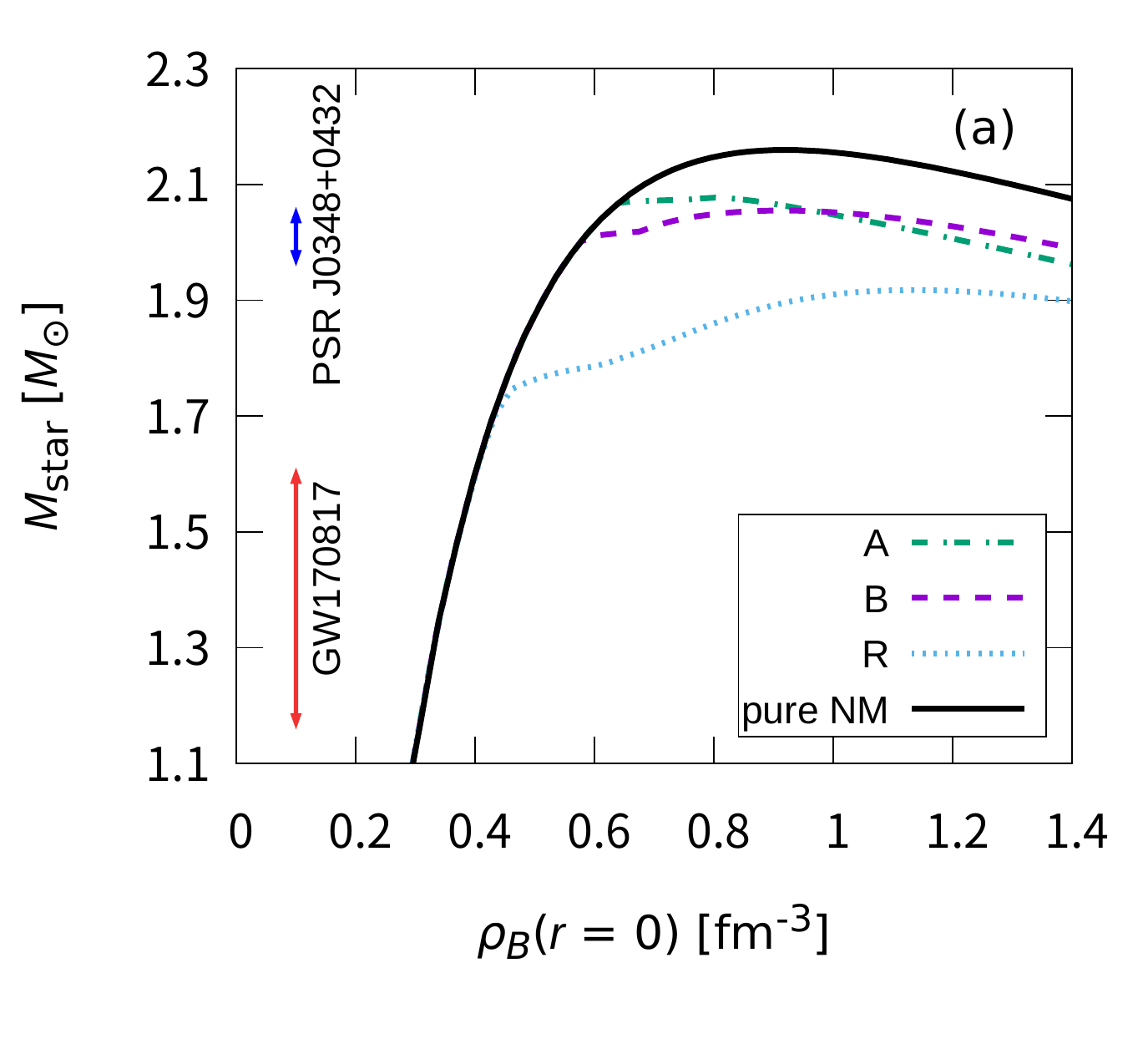} \\[1.4em]
  \centering\includegraphics[width=\columnwidth]{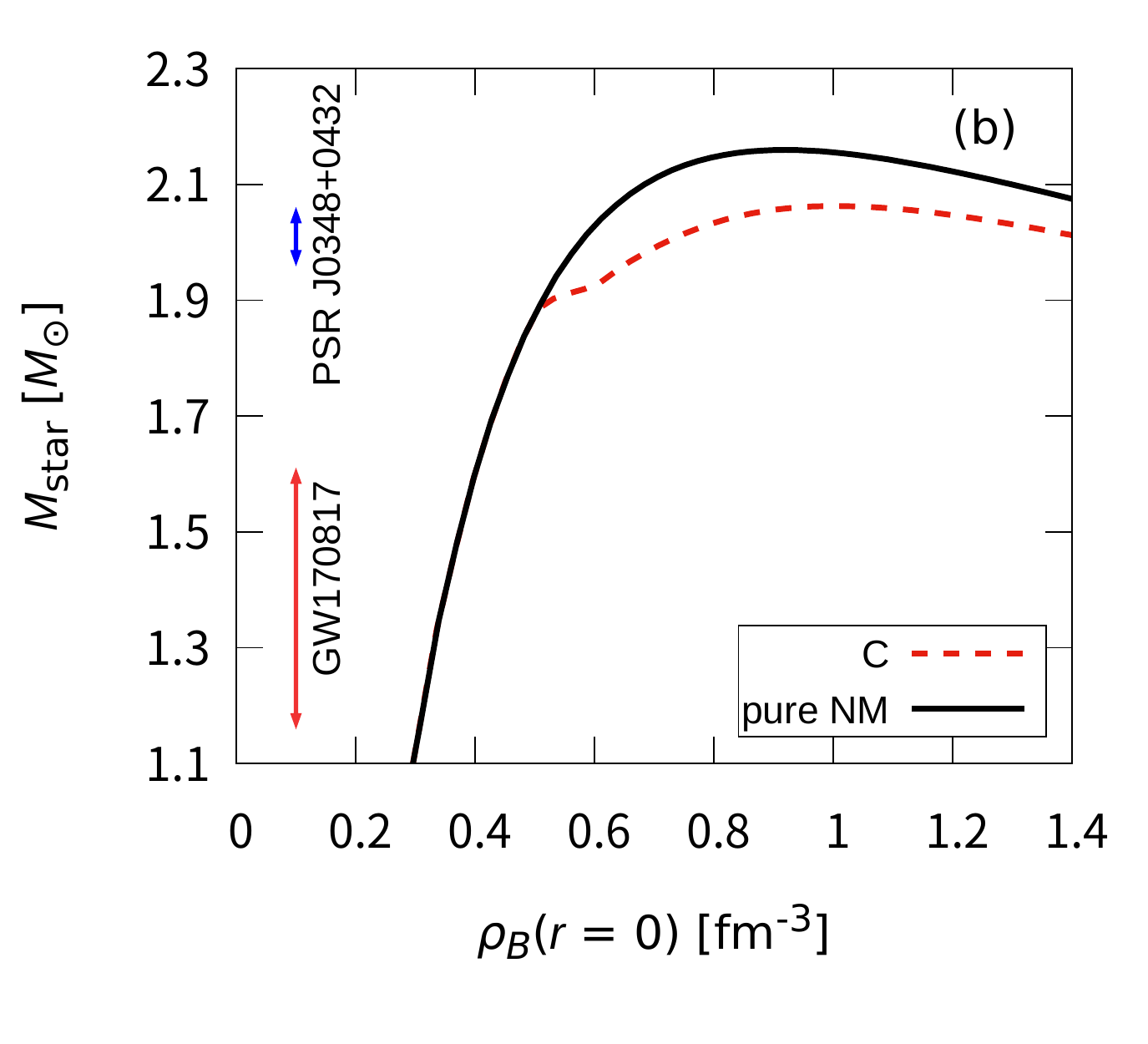}
\caption{(Color online) Neutron star masses as functions of the central baryon density $\rho_B(r=0)$. Figure (a) shows the results obtained for the points A and B in comparison to point R of Fig.~\ref{fig:map}, and the (b) shows the result obtained for the point C. The black solid line shows the pure NM case.}
\label{fig:TOV}
\end{figure}

By using our equations of state as an input to solve the Tolman-Oppenheimer-Volkoff (TOV) equations~\cite{Tolman:1939jz,Oppenheimer:1939ne}, we can calculate the properties of neutron stars. The results for the neutron star mass as a function of the central baryon density are shown in Fig.~\ref{fig:TOV}\textcolor{blue}{a} for the cases of points A, B, and R of Fig.~\ref{fig:map}, and in Fig.~\ref{fig:TOV}\textcolor{blue}{b} for case C. For case A (high transition density) one can obtain the largest star masses, but the range of stability against density fluctuations (${\rm d}M_{\rm star}/{\rm d} \rho_B(r=0) > 0$) is narrow, i.e., the hybrid star tends to be unstable. If we increase the pairing strength (A $\rightarrow$ R), the overall equation of state becomes softer and the neutron star masses become smaller, however the stability of the hybrid star against density fluctuations is substantially improved. Increasing the vector coupling (R $\rightarrow$ B), we can obtain stars which satisfy $M_{\rm star}^{\rm (max)} \geq  1.97\,M_\odot$ and show a reasonable range of stability. Fig.~\ref{fig:TOV}\textcolor{blue}{b} shows the results obtained for the point C in Fig.~\ref{fig:map}. Here the star mass near the onset of the phase transition is smaller than for the cases 
A and B, which expresses the softening of the equation of state near the phase transition density 
along the line A~$\rightarrow$~B~$\rightarrow$~C (see Fig.~\ref{fig:p-rho_B}), but for very high central densities the QM equation of state is stiff enough
to support a heavy hybrid star within a reasonably broad range of stability.  

\begin{table}[tbp]
\addtolength{\extrarowheight}{2.2pt}
    \centering
    \caption{Transition densities $\rho_\mathrm{tr}$
    and maximum star masses $M_{\rm star}^{\rm max}$ for the cases A, B, and C, in comparison
    to the case R of Fig.~\ref{fig:map}.}
    \begin{tabular*}{\columnwidth}{@{\extracolsep{\fill}}ccccc}
        \hline\hline
        Case & $c_s$ & $c_v$ & $\rho_\mathrm{tr}\,[\mathrm{fm^{-3}}]$
        & $M_\mathrm{star}^{\rm max}\,[M_\mathrm{\odot}]$
        \\[0.2em] \hline
        $\mathrm{A}$ & 0.80 & 0.50 & 0.643 & 2.078 \\
        $\mathrm{B}$ & 0.90 & 0.60 & 0.584 & 2.055 \\
        $\mathrm{C}$ & 0.98 & 0.68 & 0.496 & 2.071 \\
        \hline
        $\mathrm{R}$ & 0.50 & 0.90 & 0.379 & 1.918 \\
        \hline\hline
    \end{tabular*}
    \label{tab:map_point}
\end{table}

In Figs.~\ref{fig:TOV} we also indicate the recently observed values of neutron star masses. GW170817 denotes the neutron star coalescence event observed by the gravitational wave measurements~\cite{Abbott:2018exr,TheLIGOScientific:2017qsa,Evans:2017mmy}, where neutron stars with masses in the range $M_{\rm star} = 1.17 \sim 1.60\,M_{\odot}$ were observed and PSR J0348+0432 refers to the observation of massive neutron stars (pulsars) of mass $M_{\rm star} = 2.01 \pm 0.04\,M_{\odot}$~\cite{Antoniadis:2013pzd}. The result of PSR can be considered as a lower limit for calculations of the maximum mass of neutron stars. We see that our parameter sets A, B, and C satisfy this constraint, but the maximum star mass for case R is too small. In Tab.~\ref{tab:map_point} we list the transition densities and maximum neutron star masses for the cases A, B, and C, in comparison to the reference case R.

\begin{figure}[tbp]
  \centering\includegraphics[width=\columnwidth]{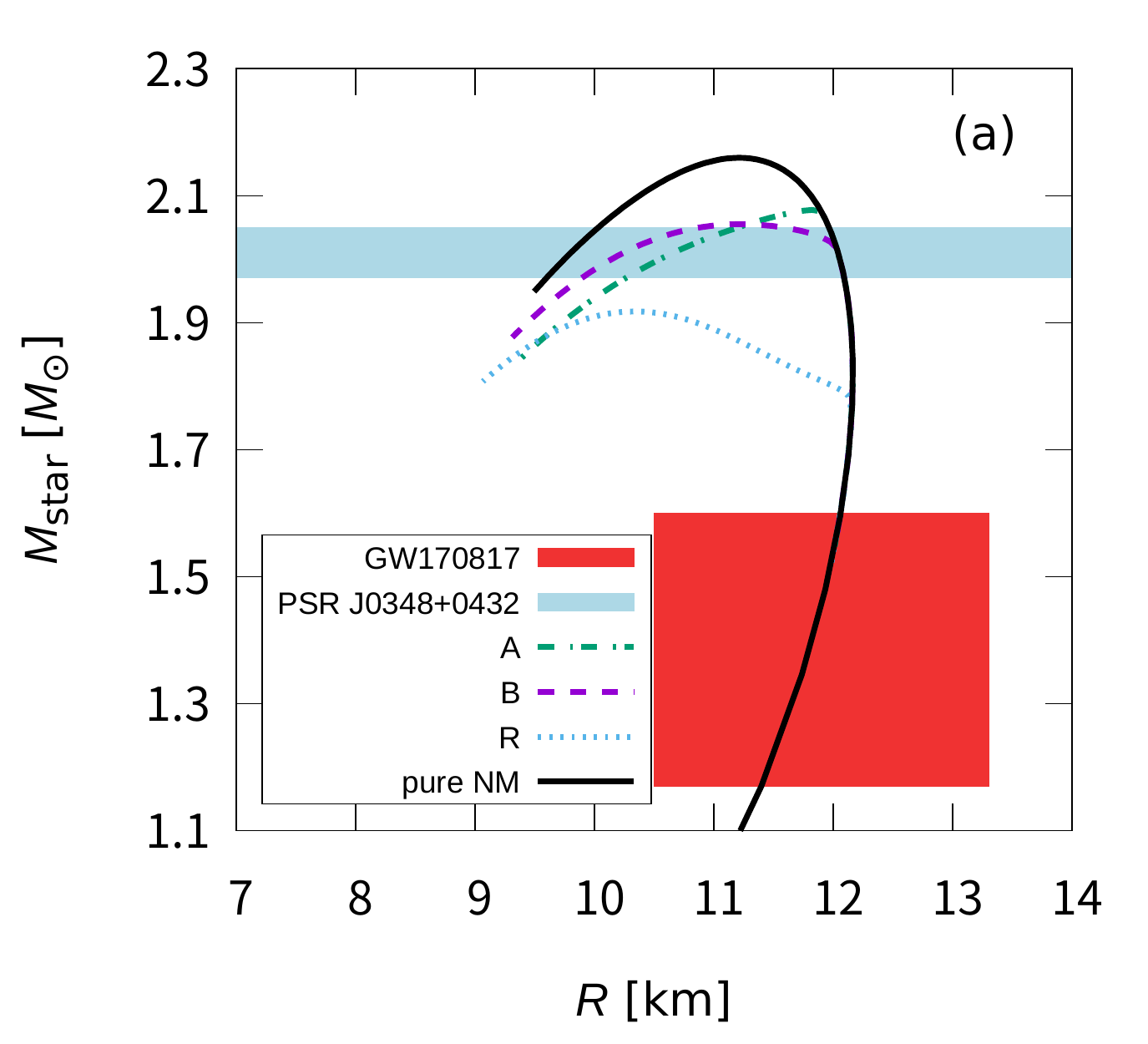} \\[0.0em]
  \centering\includegraphics[width=\columnwidth]{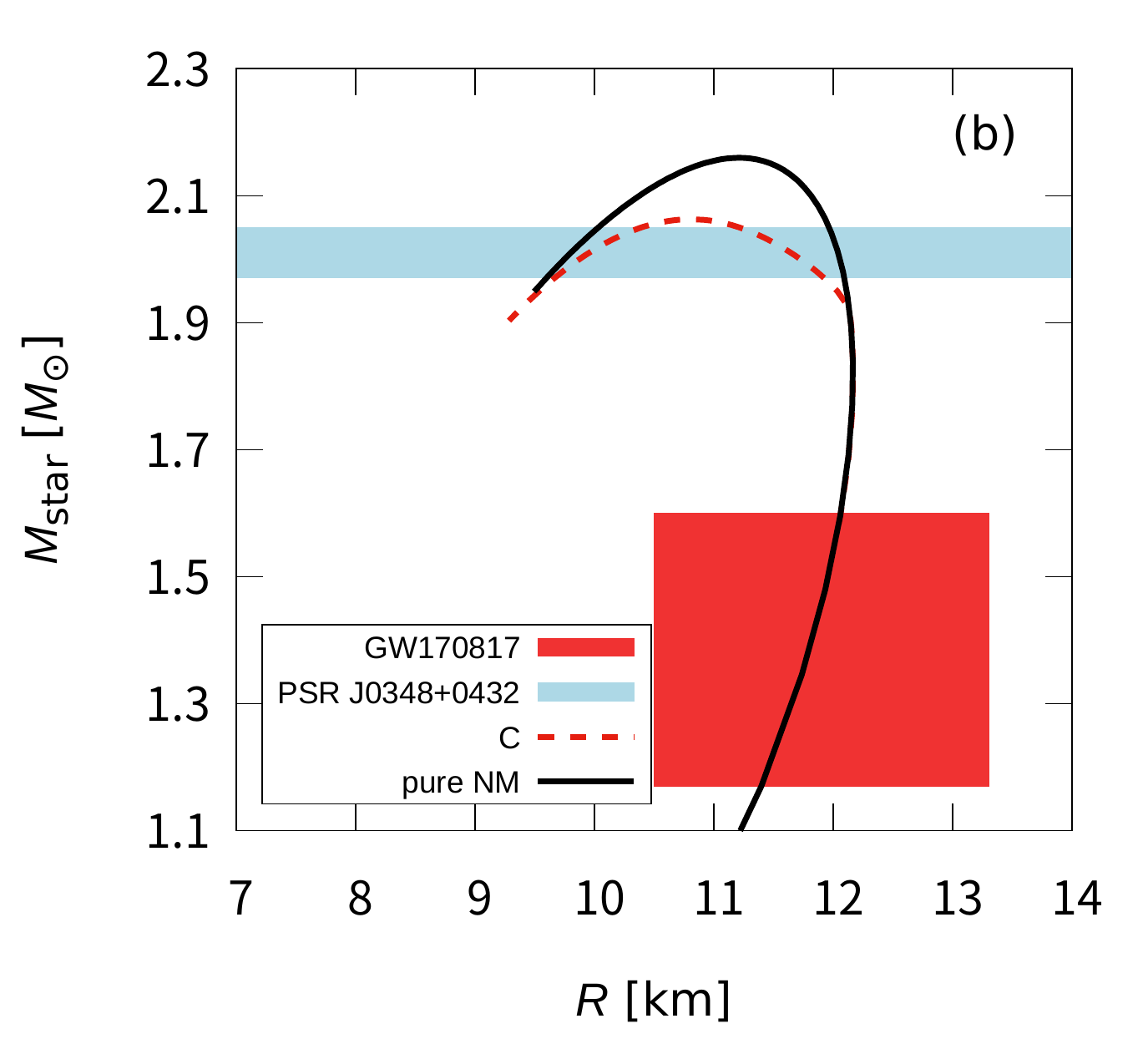}
\caption{(Color online) Mass--radius relation for neutron stars. Figure (a) shows the results obtained for the points A and B in comparison to point R of Fig.~\ref{fig:map}, and (b) shows the result obtained for the point C. The black solid line shows the pure NM case.}
\label{fig:MR}
\end{figure}

In Fig.~\ref{fig:MR}\textcolor{blue}{a} we show the relation between the neutron star masses and radii for the cases A, B, and R of Fig.~\ref{fig:map}, and Fig.~\ref{fig:MR}\textcolor{blue}{b} shows the results for the case C. The low density part of the NM curve (lower part of the solid line in Figs.~\ref{fig:MR}) shows the characteristics of a case where the pressure drops to zero (or nearly to zero) at a finite value of the baryon density, as the density decreases,\footnote{If the pressure drops to zero at $\rho_B=0$, the lower part of the NM curves in Figs.~\ref{fig:MR} would turn to the right side, instead of turning to the left.} which is indicative of a bound state in the absence of gravity. Fig.~\ref{fig:MR}\textcolor{blue}{a} clearly shows that QM in case A (highest value of $\rho_{\rm tr}$) tends to be unstable. (The narrow region of stability shown in Fig.~\ref{fig:TOV}\textcolor{blue}{a} for case A becomes invisible in Fig.~\ref{fig:MR}\textcolor{blue}{a}.) 
Similar to the previous figures, the effects of the softening of the QM equation of state
caused by increasing the pairing strength (A~$\rightarrow$~R) and the improvement of the
stability caused by increasing the vector couplings (R~$\rightarrow$~B) are clearly seen in Fig.~\ref{fig:MR}\textcolor{blue}{a}. 
The behavior of the curves with increasing vector couplings (R~$\rightarrow$~B) are
qualitatively similar to the results shown in Refs.~\cite{Contrera:2014eza,Wu:2017xaz}, 
although in those works
only small vector couplings could be used because the counteracting effects of
pairing were not taken into account. The overall softening of the QM equation of state
and the resulting improvement of the stability along the line A~$\rightarrow$~B 
of Fig.~\ref{fig:map} continues
further as we go on to the case C. In this case, the
star mass at the onset of the phase transition is lower than in cases A and B, but
the QM equation of state is stiff enough to support heavy hybrid stars with radii
of about 11\,km.   
Our results for all cases shown in Figs.~\ref{fig:MR} follow qualitatively
the ``connected'' topology of Fig. 2c in Ref.~\cite{Alford:2013aca}, i.e., the
pure neutron star and hybrid star configurations form a connected sequence.  
(As already mentioned, for case A this is not immediately evident on the scale of Fig.~\ref{fig:MR}\textcolor{blue}{a}.)
A connected sequence of this type has also been obtained in the calculations of Ref.~\cite{Abgaryan:2018gqp}, see their Fig. 7 for the case of $\Delta_P > 5 \%$. 

We finally remark that our scenario C is completely consistent with the PSR observation of massive neutron stars~\cite{Demorest:2010bx,Antoniadis:2013pzd}, as shown in Fig.~\ref{fig:MR}\textcolor{blue}{b}. The data for the event observed by GW170817~\cite{Abbott:2018exr,TheLIGOScientific:2017qsa,Evans:2017mmy} ($M_{\rm star} = 1.17 \sim 1.60\,M_{\odot}$, $R = 11.9 \pm 1.4\,$km) is also indicated in Figs.~\ref{fig:MR}, and can be reproduced with a pure NM equation of state. Before drawing firm conclusions,
however, it is most important to closely investigate the role of strangeness, both in the NM and the QM phase.



\section{SUMMARY\label{sec:summary}}
We have studied the equation of state for cold high density neutron star matter by using the two-flavor NJL model. 
Our principal aim was to use a model framework which is based on a reasonable description of hadron structure in vacuum and nuclear matter. We used the Gibbs conditions to construct the hadron-quark phase transition. Our emphasis was on the important counteracting roles played by the attractive pairing interaction (coupling constant $G_S$) and
repulsive vector interaction (coupling constants $G_{\omega}$, $G_{\rho}$) in quark matter. For this purpose, we scaled
the value of $G_S$ adjusted to the free nucleon mass by a factor $c_s$, and $G_{\omega}$, $G_{\rho}$ adjusted to the
binding energy and symmetry energy at the saturation point of isospin symmetric nuclear matter 
by a common
factor $c_v$, and investigated the dependence of the results on the parameters $(c_s, c_v)$.
We found that there exists a narrow region of ``allowed'' values of these parameters, shown in yellow in Fig.~\ref{fig:map}, which give a reasonable description of the hadron-quark phase transition and the properties of
neutron stars. Importantly, this region extends up to $c_s = 0.98$, i.e., we found that  
practically the same strength of the scalar diquark interaction, which is required to reproduce the nucleon mass and other quark-quark correlation effects in baryons~\cite{Cloet:2014rja}, can also be used to describe 
the phase transition to color superconducting quark matter and the stable massive hybrid stars. 
Concerning the vector interaction, we found that the coupling constants must be decreased in the quark matter sector by a factor of $0.45\leq c_v \leq 0.68$ to obtain a reasonable scenario for the phase transition and neutron stars. Nevertheless, we found that the inclusion of the vector repulsion in quark matter is very important, which is consistent
with earlier reported results~\cite{Baym:2017whm,Whittenbury:2015ziz,Hell:2012da}. 

Our equation of state above the hadron-quark transition density is close to a phase where chiral symmetry
is largely restored and color symmetry is strongly broken, which is consistent with the findings of
many previous works based on the 2-flavor picture of quark pairing. 
We compared our resulting equation of state with the phenomenological parametrizations of Refs.~\cite{Alford:2013aca,Ranea-Sandoval:2015ldr}, which are based
on generic stability criteria for hybrid stars, and found a qualitative and quantitative consistency for the
region of our allowed parameters. We found that the stability of massive hybrid stars favors the
largest possible values of the pairing strength and the vector couplings in quark matter.\footnote{Interestingly, a similar
preference of large coupling constants has also been found in a different treatment of the phase transition based on
hadron-quark continuity\cite{Baym:2017whm}.}      
For the case of point C in Fig.~\ref{fig:map}, we found a phase transition to QM at about three times the normal nuclear matter density, and a connected sequence of pure neutron star and hybrid star configurations.  
The maximum star mass in this case exceeds two solar masses, which is in agreement with recent observations of a massive neutron star~\cite{Demorest:2010bx,Antoniadis:2013pzd}. Our equation of state for quark matter, obtained by using the
allowed region of parameters, is stiff enough to support a heavy hybrid star within a reasonably broad range of stability.      

Although we used the Gibbs conditions to describe the hadron-quark phase transition, our results for the
allowed range of parameters are very similar to the usual Maxwellian first-order phase transition: The
variation of the pressure during the phase transition is small, and therefore the 
spatial extension of the mixed phase inside the star is small compared to its overall size, resulting
in an almost sharp boundary between the quark matter core and the surrounding nuclear matter phase.
This scenario suggests that the inclusion of finite size effects in the mixed phase, which are known to work
against spatially extended mixed phases~\cite{Endo:2011em,Neumann:2002jm,Voskresensky:2002hu,Wu:2017xaz}, 
will not lead to qualitative changes of the overall physical picture. Quantitatively, however, finite
size effects should be investigated in future applications of our model.

The calculations presented in this work should be extended to include the effects of strangeness in both the hadronic and the quark matter phases. For the case of single baryons, recent studies have shown that the properties of hyperons (masses, magnetic moments, and form factors) can be described in this NJL framework~\cite{Carrillo-Serrano:2016igi}. In the quark matter phase different types of pairings, such as color-flavor locking, should be included, and the role of a chemical potential associated with color neutrality should be taken into account. In our view this work is an important step towards a unified description of single hadrons, nuclear systems, quark matter, and neutron stars in the framework of an effective quark theory of QCD.

\begin{acknowledgments}
T.T. wishes to thank the staff and students of the Department of Physics at Tokai University for their discussions and advice. W.B. acknowledges the hospitality of Argonne National Laboratory where part of this work was performed. The work of I.C. was supported by the U.S. Department of Energy, Office of Science, Office of Nuclear Physics, contract no. DE-AC02-06CH11357.
\end{acknowledgments}


\begin{thebibliography}{84}%
\makeatletter
\providecommand \@ifxundefined [1]{%
 \@ifx{#1\undefined}
}%
\providecommand \@ifnum [1]{%
 \ifnum #1\expandafter \@firstoftwo
 \else \expandafter \@secondoftwo
 \fi
}%
\providecommand \@ifx [1]{%
 \ifx #1\expandafter \@firstoftwo
 \else \expandafter \@secondoftwo
 \fi
}%
\providecommand \natexlab [1]{#1}%
\providecommand \enquote  [1]{``#1''}%
\providecommand \bibnamefont  [1]{#1}%
\providecommand \bibfnamefont [1]{#1}%
\providecommand \citenamefont [1]{#1}%
\providecommand \href@noop [0]{\@secondoftwo}%
\providecommand \href [0]{\begingroup \@sanitize@url \@href}%
\providecommand \@href[1]{\@@startlink{#1}\@@href}%
\providecommand \@@href[1]{\endgroup#1\@@endlink}%
\providecommand \@sanitize@url [0]{\catcode `\\12\catcode `\$12\catcode
  `\&12\catcode `\#12\catcode `\^12\catcode `\_12\catcode `\%12\relax}%
\providecommand \@@startlink[1]{}%
\providecommand \@@endlink[0]{}%
\providecommand \url  [0]{\begingroup\@sanitize@url \@url }%
\providecommand \@url [1]{\endgroup\@href {#1}{\urlprefix }}%
\providecommand \urlprefix  [0]{URL }%
\providecommand \Eprint [0]{\href }%
\providecommand \doibase [0]{http://dx.doi.org/}%
\providecommand \selectlanguage [0]{\@gobble}%
\providecommand \bibinfo  [0]{\@secondoftwo}%
\providecommand \bibfield  [0]{\@secondoftwo}%
\providecommand \translation [1]{[#1]}%
\providecommand \BibitemOpen [0]{}%
\providecommand \bibitemStop [0]{}%
\providecommand \bibitemNoStop [0]{.\EOS\space}%
\providecommand \EOS [0]{\spacefactor3000\relax}%
\providecommand \BibitemShut  [1]{\csname bibitem#1\endcsname}%
\let\auto@bib@innerbib\@empty
\bibitem [{\citenamefont {Itoh}(1970)}]{Itoh:1970uw}%
  \BibitemOpen
  \bibfield  {author} {\bibinfo {author} {\bibfnamefont {N.}~\bibnamefont
  {Itoh}},\ }\href {\doibase 10.1143/PTP.44.291} {\bibfield  {journal}
  {\bibinfo  {journal} {Prog. Theor. Phys.}\ }\textbf {\bibinfo {volume}
  {44}},\ \bibinfo {pages} {291} (\bibinfo {year} {1970})}\BibitemShut
  {NoStop}%
\bibitem [{\citenamefont {Baym}\ \emph {et~al.}(1971)\citenamefont {Baym},
  \citenamefont {Pethick},\ and\ \citenamefont {Sutherland}}]{Baym:1971pw}%
  \BibitemOpen
  \bibfield  {author} {\bibinfo {author} {\bibfnamefont {G.}~\bibnamefont
  {Baym}}, \bibinfo {author} {\bibfnamefont {C.}~\bibnamefont {Pethick}}, \
  and\ \bibinfo {author} {\bibfnamefont {P.}~\bibnamefont {Sutherland}},\
  }\href {\doibase 10.1086/151216} {\bibfield  {journal} {\bibinfo  {journal}
  {Astrophys. J.}\ }\textbf {\bibinfo {volume} {170}},\ \bibinfo {pages} {299}
  (\bibinfo {year} {1971})}\BibitemShut {NoStop}%
\bibitem [{\citenamefont {Glendenning}(1997)}]{Glendenning:1997wn}%
  \BibitemOpen
  \bibfield  {author} {\bibinfo {author} {\bibfnamefont {N.~K.}\ \bibnamefont
  {Glendenning}},\ }\href@noop {} {\emph {\bibinfo {title} {{Compact stars:
  Nuclear physics, particle physics, and general relativity}}}}\ (\bibinfo
  {year} {1997})\BibitemShut {NoStop}%
\bibitem [{\citenamefont {Heiselberg}\ and\ \citenamefont
  {Hjorth-Jensen}(2000)}]{Heiselberg:1999mq}%
  \BibitemOpen
  \bibfield  {author} {\bibinfo {author} {\bibfnamefont {H.}~\bibnamefont
  {Heiselberg}}\ and\ \bibinfo {author} {\bibfnamefont {M.}~\bibnamefont
  {Hjorth-Jensen}},\ }\href {\doibase 10.1016/S0370-1573(99)00110-6} {\bibfield
   {journal} {\bibinfo  {journal} {Phys. Rept.}\ }\textbf {\bibinfo {volume}
  {328}},\ \bibinfo {pages} {237} (\bibinfo {year} {2000})},\ \Eprint
  {http://arxiv.org/abs/nucl-th/9902033} {arXiv:nucl-th/9902033 [nucl-th]}
  \BibitemShut {NoStop}%
\bibitem [{\citenamefont {Dean}\ and\ \citenamefont
  {Hjorth-Jensen}(2003)}]{Dean:2002zx}%
  \BibitemOpen
  \bibfield  {author} {\bibinfo {author} {\bibfnamefont {D.~J.}\ \bibnamefont
  {Dean}}\ and\ \bibinfo {author} {\bibfnamefont {M.}~\bibnamefont
  {Hjorth-Jensen}},\ }\href {\doibase 10.1103/RevModPhys.75.607} {\bibfield
  {journal} {\bibinfo  {journal} {Rev. Mod. Phys.}\ }\textbf {\bibinfo {volume}
  {75}},\ \bibinfo {pages} {607} (\bibinfo {year} {2003})},\ \Eprint
  {http://arxiv.org/abs/nucl-th/0210033} {arXiv:nucl-th/0210033 [nucl-th]}
  \BibitemShut {NoStop}%
\bibitem [{\citenamefont {Baym}\ \emph {et~al.}(2018)\citenamefont {Baym},
  \citenamefont {Hatsuda}, \citenamefont {Kojo}, \citenamefont {Powell},
  \citenamefont {Song},\ and\ \citenamefont {Takatsuka}}]{Baym:2017whm}%
  \BibitemOpen
  \bibfield  {author} {\bibinfo {author} {\bibfnamefont {G.}~\bibnamefont
  {Baym}}, \bibinfo {author} {\bibfnamefont {T.}~\bibnamefont {Hatsuda}},
  \bibinfo {author} {\bibfnamefont {T.}~\bibnamefont {Kojo}}, \bibinfo {author}
  {\bibfnamefont {P.~D.}\ \bibnamefont {Powell}}, \bibinfo {author}
  {\bibfnamefont {Y.}~\bibnamefont {Song}}, \ and\ \bibinfo {author}
  {\bibfnamefont {T.}~\bibnamefont {Takatsuka}},\ }\href {\doibase
  10.1088/1361-6633/aaae14} {\bibfield  {journal} {\bibinfo  {journal} {Rept.
  Prog. Phys.}\ }\textbf {\bibinfo {volume} {81}},\ \bibinfo {pages} {056902}
  (\bibinfo {year} {2018})},\ \Eprint {http://arxiv.org/abs/1707.04966}
  {arXiv:1707.04966 [astro-ph.HE]} \BibitemShut {NoStop}%
\bibitem [{\citenamefont {Demorest}\ \emph {et~al.}(2010)\citenamefont
  {Demorest}, \citenamefont {Pennucci}, \citenamefont {Ransom}, \citenamefont
  {Roberts},\ and\ \citenamefont {Hessels}}]{Demorest:2010bx}%
  \BibitemOpen
  \bibfield  {author} {\bibinfo {author} {\bibfnamefont {P.}~\bibnamefont
  {Demorest}}, \bibinfo {author} {\bibfnamefont {T.}~\bibnamefont {Pennucci}},
  \bibinfo {author} {\bibfnamefont {S.}~\bibnamefont {Ransom}}, \bibinfo
  {author} {\bibfnamefont {M.}~\bibnamefont {Roberts}}, \ and\ \bibinfo
  {author} {\bibfnamefont {J.}~\bibnamefont {Hessels}},\ }\href {\doibase
  10.1038/nature09466} {\bibfield  {journal} {\bibinfo  {journal} {Nature}\
  }\textbf {\bibinfo {volume} {467}},\ \bibinfo {pages} {1081} (\bibinfo {year}
  {2010})},\ \Eprint {http://arxiv.org/abs/1010.5788} {arXiv:1010.5788
  [astro-ph.HE]} \BibitemShut {NoStop}%
\bibitem [{\citenamefont {Antoniadis}\ \emph {et~al.}(2013)\citenamefont
  {Antoniadis} \emph {et~al.}}]{Antoniadis:2013pzd}%
  \BibitemOpen
  \bibfield  {author} {\bibinfo {author} {\bibfnamefont {J.}~\bibnamefont
  {Antoniadis}} \emph {et~al.},\ }\href {\doibase 10.1126/science.1233232}
  {\bibfield  {journal} {\bibinfo  {journal} {Science}\ }\textbf {\bibinfo
  {volume} {340}},\ \bibinfo {pages} {6131} (\bibinfo {year} {2013})},\ \Eprint
  {http://arxiv.org/abs/1304.6875} {arXiv:1304.6875 [astro-ph.HE]} \BibitemShut
  {NoStop}%
\bibitem [{\citenamefont {Abbott}\ \emph {et~al.}(2018)\citenamefont {Abbott}
  \emph {et~al.}}]{Abbott:2018exr}%
  \BibitemOpen
  \bibfield  {author} {\bibinfo {author} {\bibfnamefont {B.~P.}\ \bibnamefont
  {Abbott}} \emph {et~al.} (\bibinfo {collaboration} {LIGO Scientific,
  Virgo}),\ }\href {\doibase 10.1103/PhysRevLett.121.161101} {\bibfield
  {journal} {\bibinfo  {journal} {Phys. Rev. Lett.}\ }\textbf {\bibinfo
  {volume} {121}},\ \bibinfo {pages} {161101} (\bibinfo {year} {2018})},\
  \Eprint {http://arxiv.org/abs/1805.11581} {arXiv:1805.11581 [gr-qc]}
  \BibitemShut {NoStop}%
\bibitem [{\citenamefont {Abbott}\ \emph {et~al.}(2017)\citenamefont {Abbott}
  \emph {et~al.}}]{TheLIGOScientific:2017qsa}%
  \BibitemOpen
  \bibfield  {author} {\bibinfo {author} {\bibfnamefont {B.}~\bibnamefont
  {Abbott}} \emph {et~al.} (\bibinfo {collaboration} {LIGO Scientific,
  Virgo}),\ }\href {\doibase 10.1103/PhysRevLett.119.161101} {\bibfield
  {journal} {\bibinfo  {journal} {Phys. Rev. Lett.}\ }\textbf {\bibinfo
  {volume} {119}},\ \bibinfo {pages} {161101} (\bibinfo {year} {2017})},\
  \Eprint {http://arxiv.org/abs/1710.05832} {arXiv:1710.05832 [gr-qc]}
  \BibitemShut {NoStop}%
\bibitem [{\citenamefont {Evans}\ \emph {et~al.}(2017)\citenamefont {Evans}
  \emph {et~al.}}]{Evans:2017mmy}%
  \BibitemOpen
  \bibfield  {author} {\bibinfo {author} {\bibfnamefont {P.~A.}\ \bibnamefont
  {Evans}} \emph {et~al.},\ }\href {\doibase 10.1126/science.aap9580}
  {\bibfield  {journal} {\bibinfo  {journal} {Science}\ }\textbf {\bibinfo
  {volume} {358}},\ \bibinfo {pages} {1565} (\bibinfo {year} {2017})},\ \Eprint
  {http://arxiv.org/abs/1710.05437} {arXiv:1710.05437 [astro-ph.HE]}
  \BibitemShut {NoStop}%
\bibitem [{\citenamefont {Gandolfi}\ \emph {et~al.}(2012)\citenamefont
  {Gandolfi}, \citenamefont {Carlson},\ and\ \citenamefont
  {Reddy}}]{Gandolfi:2011xu}%
  \BibitemOpen
  \bibfield  {author} {\bibinfo {author} {\bibfnamefont {S.}~\bibnamefont
  {Gandolfi}}, \bibinfo {author} {\bibfnamefont {J.}~\bibnamefont {Carlson}}, \
  and\ \bibinfo {author} {\bibfnamefont {S.}~\bibnamefont {Reddy}},\ }\href
  {\doibase 10.1103/PhysRevC.85.032801} {\bibfield  {journal} {\bibinfo
  {journal} {Phys. Rev.}\ }\textbf {\bibinfo {volume} {C85}},\ \bibinfo {pages}
  {032801} (\bibinfo {year} {2012})},\ \Eprint {http://arxiv.org/abs/1101.1921}
  {arXiv:1101.1921 [nucl-th]} \BibitemShut {NoStop}%
\bibitem [{\citenamefont {Gandolfi}\ \emph {et~al.}(2014)\citenamefont
  {Gandolfi}, \citenamefont {Carlson}, \citenamefont {Reddy}, \citenamefont
  {Steiner},\ and\ \citenamefont {Wiringa}}]{Gandolfi:2013baa}%
  \BibitemOpen
  \bibfield  {author} {\bibinfo {author} {\bibfnamefont {S.}~\bibnamefont
  {Gandolfi}}, \bibinfo {author} {\bibfnamefont {J.}~\bibnamefont {Carlson}},
  \bibinfo {author} {\bibfnamefont {S.}~\bibnamefont {Reddy}}, \bibinfo
  {author} {\bibfnamefont {A.~W.}\ \bibnamefont {Steiner}}, \ and\ \bibinfo
  {author} {\bibfnamefont {R.~B.}\ \bibnamefont {Wiringa}},\ }\href {\doibase
  10.1140/epja/i2014-14010-5} {\bibfield  {journal} {\bibinfo  {journal} {Eur.
  Phys. J.}\ }\textbf {\bibinfo {volume} {A50}},\ \bibinfo {pages} {10}
  (\bibinfo {year} {2014})},\ \Eprint {http://arxiv.org/abs/1307.5815}
  {arXiv:1307.5815 [nucl-th]} \BibitemShut {NoStop}%
\bibitem [{\citenamefont {Hebeler}\ \emph {et~al.}(2011)\citenamefont
  {Hebeler}, \citenamefont {Bogner}, \citenamefont {Furnstahl}, \citenamefont
  {Nogga},\ and\ \citenamefont {Schwenk}}]{Hebeler:2010xb}%
  \BibitemOpen
  \bibfield  {author} {\bibinfo {author} {\bibfnamefont {K.}~\bibnamefont
  {Hebeler}}, \bibinfo {author} {\bibfnamefont {S.~K.}\ \bibnamefont {Bogner}},
  \bibinfo {author} {\bibfnamefont {R.~J.}\ \bibnamefont {Furnstahl}}, \bibinfo
  {author} {\bibfnamefont {A.}~\bibnamefont {Nogga}}, \ and\ \bibinfo {author}
  {\bibfnamefont {A.}~\bibnamefont {Schwenk}},\ }\href {\doibase
  10.1103/PhysRevC.83.031301} {\bibfield  {journal} {\bibinfo  {journal} {Phys.
  Rev.}\ }\textbf {\bibinfo {volume} {C83}},\ \bibinfo {pages} {031301}
  (\bibinfo {year} {2011})},\ \Eprint {http://arxiv.org/abs/1012.3381}
  {arXiv:1012.3381 [nucl-th]} \BibitemShut {NoStop}%
\bibitem [{\citenamefont {Kaiser}(2012)}]{Kaiser:2012ex}%
  \BibitemOpen
  \bibfield  {author} {\bibinfo {author} {\bibfnamefont {N.}~\bibnamefont
  {Kaiser}},\ }\href {\doibase 10.1140/epja/i2012-12135-1} {\bibfield
  {journal} {\bibinfo  {journal} {Eur. Phys. J.}\ }\textbf {\bibinfo {volume}
  {A48}},\ \bibinfo {pages} {135} (\bibinfo {year} {2012})},\ \Eprint
  {http://arxiv.org/abs/1209.4556} {arXiv:1209.4556 [nucl-th]} \BibitemShut
  {NoStop}%
\bibitem [{\citenamefont {Muller}\ and\ \citenamefont
  {Serot}(1995)}]{Muller:1995ji}%
  \BibitemOpen
  \bibfield  {author} {\bibinfo {author} {\bibfnamefont {H.}~\bibnamefont
  {Muller}}\ and\ \bibinfo {author} {\bibfnamefont {B.~D.}\ \bibnamefont
  {Serot}},\ }\href {\doibase 10.1103/PhysRevC.52.2072} {\bibfield  {journal}
  {\bibinfo  {journal} {Phys. Rev.}\ }\textbf {\bibinfo {volume} {C52}},\
  \bibinfo {pages} {2072} (\bibinfo {year} {1995})},\ \Eprint
  {http://arxiv.org/abs/nucl-th/9505013} {arXiv:nucl-th/9505013 [nucl-th]}
  \BibitemShut {NoStop}%
\bibitem [{\citenamefont {Chin}\ and\ \citenamefont
  {Walecka}(1974)}]{Chin:1974sa}%
  \BibitemOpen
  \bibfield  {author} {\bibinfo {author} {\bibfnamefont {S.~A.}\ \bibnamefont
  {Chin}}\ and\ \bibinfo {author} {\bibfnamefont {J.~D.}\ \bibnamefont
  {Walecka}},\ }\href {\doibase 10.1016/0370-2693(74)90708-4} {\bibfield
  {journal} {\bibinfo  {journal} {Phys. Lett.}\ }\textbf {\bibinfo {volume}
  {52B}},\ \bibinfo {pages} {24} (\bibinfo {year} {1974})}\BibitemShut
  {NoStop}%
\bibitem [{\citenamefont {Kaplan}\ and\ \citenamefont
  {Nelson}(1986)}]{Kaplan:1986yq}%
  \BibitemOpen
  \bibfield  {author} {\bibinfo {author} {\bibfnamefont {D.~B.}\ \bibnamefont
  {Kaplan}}\ and\ \bibinfo {author} {\bibfnamefont {A.~E.}\ \bibnamefont
  {Nelson}},\ }\href {\doibase 10.1016/0370-2693(86)90331-X} {\bibfield
  {journal} {\bibinfo  {journal} {Phys. Lett.}\ }\textbf {\bibinfo {volume}
  {B175}},\ \bibinfo {pages} {57} (\bibinfo {year} {1986})}\BibitemShut
  {NoStop}%
\bibitem [{\citenamefont {Dapo}\ \emph {et~al.}(2008)\citenamefont {Dapo},
  \citenamefont {Schaefer},\ and\ \citenamefont {Wambach}}]{Dapo:2008qv}%
  \BibitemOpen
  \bibfield  {author} {\bibinfo {author} {\bibfnamefont {H.}~\bibnamefont
  {Dapo}}, \bibinfo {author} {\bibfnamefont {B.-J.}\ \bibnamefont {Schaefer}},
  \ and\ \bibinfo {author} {\bibfnamefont {J.}~\bibnamefont {Wambach}},\ }\href
  {\doibase 10.1140/epja/i2008-10542-5} {\bibfield  {journal} {\bibinfo
  {journal} {Eur. Phys. J.}\ }\textbf {\bibinfo {volume} {A36}},\ \bibinfo
  {pages} {101} (\bibinfo {year} {2008})},\ \Eprint
  {http://arxiv.org/abs/0802.2646} {arXiv:0802.2646 [nucl-th]} \BibitemShut
  {NoStop}%
\bibitem [{\citenamefont {Nambu}\ and\ \citenamefont
  {Jona-Lasinio}(1961{\natexlab{a}})}]{Nambu:1961tp}%
  \BibitemOpen
  \bibfield  {author} {\bibinfo {author} {\bibfnamefont {Y.}~\bibnamefont
  {Nambu}}\ and\ \bibinfo {author} {\bibfnamefont {G.}~\bibnamefont
  {Jona-Lasinio}},\ }\href {\doibase 10.1103/PhysRev.122.345} {\bibfield
  {journal} {\bibinfo  {journal} {Phys. Rev.}\ }\textbf {\bibinfo {volume}
  {122}},\ \bibinfo {pages} {345} (\bibinfo {year} {1961}{\natexlab{a}})},\
  \bibinfo {note} {[,127(1961)]}\BibitemShut {NoStop}%
\bibitem [{\citenamefont {Nambu}\ and\ \citenamefont
  {Jona-Lasinio}(1961{\natexlab{b}})}]{Nambu:1961fr}%
  \BibitemOpen
  \bibfield  {author} {\bibinfo {author} {\bibfnamefont {Y.}~\bibnamefont
  {Nambu}}\ and\ \bibinfo {author} {\bibfnamefont {G.}~\bibnamefont
  {Jona-Lasinio}},\ }\href {\doibase 10.1103/PhysRev.124.246} {\bibfield
  {journal} {\bibinfo  {journal} {Phys. Rev.}\ }\textbf {\bibinfo {volume}
  {124}},\ \bibinfo {pages} {246} (\bibinfo {year} {1961}{\natexlab{b}})},\
  \bibinfo {note} {[,141(1961)]}\BibitemShut {NoStop}%
\bibitem [{\citenamefont {Vogl}\ and\ \citenamefont
  {Weise}(1991)}]{Vogl:1991qt}%
  \BibitemOpen
  \bibfield  {author} {\bibinfo {author} {\bibfnamefont {U.}~\bibnamefont
  {Vogl}}\ and\ \bibinfo {author} {\bibfnamefont {W.}~\bibnamefont {Weise}},\
  }\href {\doibase 10.1016/0146-6410(91)90005-9} {\bibfield  {journal}
  {\bibinfo  {journal} {Prog. Part. Nucl. Phys.}\ }\textbf {\bibinfo {volume}
  {27}},\ \bibinfo {pages} {195} (\bibinfo {year} {1991})}\BibitemShut
  {NoStop}%
\bibitem [{\citenamefont {Hatsuda}\ and\ \citenamefont
  {Kunihiro}(1994)}]{Hatsuda:1994pi}%
  \BibitemOpen
  \bibfield  {author} {\bibinfo {author} {\bibfnamefont {T.}~\bibnamefont
  {Hatsuda}}\ and\ \bibinfo {author} {\bibfnamefont {T.}~\bibnamefont
  {Kunihiro}},\ }\href {\doibase 10.1016/0370-1573(94)90022-1} {\bibfield
  {journal} {\bibinfo  {journal} {Phys. Rept.}\ }\textbf {\bibinfo {volume}
  {247}},\ \bibinfo {pages} {221} (\bibinfo {year} {1994})},\ \Eprint
  {http://arxiv.org/abs/hep-ph/9401310} {arXiv:hep-ph/9401310 [hep-ph]}
  \BibitemShut {NoStop}%
\bibitem [{\citenamefont {Buballa}(2005)}]{Buballa:2003qv}%
  \BibitemOpen
  \bibfield  {author} {\bibinfo {author} {\bibfnamefont {M.}~\bibnamefont
  {Buballa}},\ }\href {\doibase 10.1016/j.physrep.2004.11.004} {\bibfield
  {journal} {\bibinfo  {journal} {Phys. Rept.}\ }\textbf {\bibinfo {volume}
  {407}},\ \bibinfo {pages} {205} (\bibinfo {year} {2005})},\ \Eprint
  {http://arxiv.org/abs/hep-ph/0402234} {arXiv:hep-ph/0402234 [hep-ph]}
  \BibitemShut {NoStop}%
\bibitem [{\citenamefont {Shovkovy}(2005)}]{Shovkovy:2004me}%
  \BibitemOpen
  \bibfield  {author} {\bibinfo {author} {\bibfnamefont {I.~A.}\ \bibnamefont
  {Shovkovy}},\ }\bibfield  {booktitle} {\emph {\bibinfo {booktitle}
  {{Proceedings, 4th Biennial Conference on Classical and Quantum Relativistic
  Dynamics of Particles and Fields (IARD 2004): Saas Fee, Switzerland, June
  12-19, 2004}}},\ }\href {\doibase 10.1007/s10701-005-6440-x} {\bibfield
  {journal} {\bibinfo  {journal} {Found. Phys.}\ }\textbf {\bibinfo {volume}
  {35}},\ \bibinfo {pages} {1309} (\bibinfo {year} {2005})},\ \bibinfo {note}
  {[,260(2004)]},\ \Eprint {http://arxiv.org/abs/nucl-th/0410091}
  {arXiv:nucl-th/0410091 [nucl-th]} \BibitemShut {NoStop}%
\bibitem [{\citenamefont {Alford}\ \emph {et~al.}(2008)\citenamefont {Alford},
  \citenamefont {Schmitt}, \citenamefont {Rajagopal},\ and\ \citenamefont
  {Schäfer}}]{Alford:2007xm}%
  \BibitemOpen
  \bibfield  {author} {\bibinfo {author} {\bibfnamefont {M.~G.}\ \bibnamefont
  {Alford}}, \bibinfo {author} {\bibfnamefont {A.}~\bibnamefont {Schmitt}},
  \bibinfo {author} {\bibfnamefont {K.}~\bibnamefont {Rajagopal}}, \ and\
  \bibinfo {author} {\bibfnamefont {T.}~\bibnamefont {Schäfer}},\ }\href
  {\doibase 10.1103/RevModPhys.80.1455} {\bibfield  {journal} {\bibinfo
  {journal} {Rev. Mod. Phys.}\ }\textbf {\bibinfo {volume} {80}},\ \bibinfo
  {pages} {1455} (\bibinfo {year} {2008})},\ \Eprint
  {http://arxiv.org/abs/0709.4635} {arXiv:0709.4635 [hep-ph]} \BibitemShut
  {NoStop}%
\bibitem [{\citenamefont {Fukushima}\ and\ \citenamefont
  {Hatsuda}(2011)}]{Fukushima:2010bq}%
  \BibitemOpen
  \bibfield  {author} {\bibinfo {author} {\bibfnamefont {K.}~\bibnamefont
  {Fukushima}}\ and\ \bibinfo {author} {\bibfnamefont {T.}~\bibnamefont
  {Hatsuda}},\ }\href {\doibase 10.1088/0034-4885/74/1/014001} {\bibfield
  {journal} {\bibinfo  {journal} {Rept. Prog. Phys.}\ }\textbf {\bibinfo
  {volume} {74}},\ \bibinfo {pages} {014001} (\bibinfo {year} {2011})},\
  \Eprint {http://arxiv.org/abs/1005.4814} {arXiv:1005.4814 [hep-ph]}
  \BibitemShut {NoStop}%
\bibitem [{\citenamefont {Bailin}\ and\ \citenamefont
  {Love}(1984)}]{Bailin:1983bm}%
  \BibitemOpen
  \bibfield  {author} {\bibinfo {author} {\bibfnamefont {D.}~\bibnamefont
  {Bailin}}\ and\ \bibinfo {author} {\bibfnamefont {A.}~\bibnamefont {Love}},\
  }\href {\doibase 10.1016/0370-1573(84)90145-5} {\bibfield  {journal}
  {\bibinfo  {journal} {Phys. Rept.}\ }\textbf {\bibinfo {volume} {107}},\
  \bibinfo {pages} {325} (\bibinfo {year} {1984})}\BibitemShut {NoStop}%
\bibitem [{\citenamefont {Alford}\ \emph {et~al.}(1998)\citenamefont {Alford},
  \citenamefont {Rajagopal},\ and\ \citenamefont {Wilczek}}]{Alford:1997zt}%
  \BibitemOpen
  \bibfield  {author} {\bibinfo {author} {\bibfnamefont {M.~G.}\ \bibnamefont
  {Alford}}, \bibinfo {author} {\bibfnamefont {K.}~\bibnamefont {Rajagopal}}, \
  and\ \bibinfo {author} {\bibfnamefont {F.}~\bibnamefont {Wilczek}},\ }\href
  {\doibase 10.1016/S0370-2693(98)00051-3} {\bibfield  {journal} {\bibinfo
  {journal} {Phys. Lett.}\ }\textbf {\bibinfo {volume} {B422}},\ \bibinfo
  {pages} {247} (\bibinfo {year} {1998})},\ \Eprint
  {http://arxiv.org/abs/hep-ph/9711395} {arXiv:hep-ph/9711395 [hep-ph]}
  \BibitemShut {NoStop}%
\bibitem [{\citenamefont {Rapp}\ \emph {et~al.}(1998)\citenamefont {Rapp},
  \citenamefont {Schäfer}, \citenamefont {Shuryak},\ and\ \citenamefont
  {Velkovsky}}]{Rapp:1997zu}%
  \BibitemOpen
  \bibfield  {author} {\bibinfo {author} {\bibfnamefont {R.}~\bibnamefont
  {Rapp}}, \bibinfo {author} {\bibfnamefont {T.}~\bibnamefont {Schäfer}},
  \bibinfo {author} {\bibfnamefont {E.~V.}\ \bibnamefont {Shuryak}}, \ and\
  \bibinfo {author} {\bibfnamefont {M.}~\bibnamefont {Velkovsky}},\ }\href
  {\doibase 10.1103/PhysRevLett.81.53} {\bibfield  {journal} {\bibinfo
  {journal} {Phys. Rev. Lett.}\ }\textbf {\bibinfo {volume} {81}},\ \bibinfo
  {pages} {53} (\bibinfo {year} {1998})},\ \Eprint
  {http://arxiv.org/abs/hep-ph/9711396} {arXiv:hep-ph/9711396 [hep-ph]}
  \BibitemShut {NoStop}%
\bibitem [{\citenamefont {Glendenning}(1992)}]{Glendenning:1992vb}%
  \BibitemOpen
  \bibfield  {author} {\bibinfo {author} {\bibfnamefont {N.~K.}\ \bibnamefont
  {Glendenning}},\ }\href {\doibase 10.1103/PhysRevD.46.1274} {\bibfield
  {journal} {\bibinfo  {journal} {Phys. Rev.}\ }\textbf {\bibinfo {volume}
  {D46}},\ \bibinfo {pages} {1274} (\bibinfo {year} {1992})}\BibitemShut
  {NoStop}%
\bibitem [{\citenamefont {Wu}\ and\ \citenamefont {Shen}(2017)}]{Wu:2017xaz}%
  \BibitemOpen
  \bibfield  {author} {\bibinfo {author} {\bibfnamefont {X.}~\bibnamefont
  {Wu}}\ and\ \bibinfo {author} {\bibfnamefont {H.}~\bibnamefont {Shen}},\
  }\href {\doibase 10.1103/PhysRevC.96.025802} {\bibfield  {journal} {\bibinfo
  {journal} {Phys. Rev.}\ }\textbf {\bibinfo {volume} {C96}},\ \bibinfo {pages}
  {025802} (\bibinfo {year} {2017})},\ \Eprint
  {http://arxiv.org/abs/1708.01878} {arXiv:1708.01878 [nucl-th]} \BibitemShut
  {NoStop}%
\bibitem [{\citenamefont {Schäfer}\ and\ \citenamefont
  {Wilczek}(1999)}]{Schafer:1998ef}%
  \BibitemOpen
  \bibfield  {author} {\bibinfo {author} {\bibfnamefont {T.}~\bibnamefont
  {Schäfer}}\ and\ \bibinfo {author} {\bibfnamefont {F.}~\bibnamefont
  {Wilczek}},\ }\href {\doibase 10.1103/PhysRevLett.82.3956} {\bibfield
  {journal} {\bibinfo  {journal} {Phys. Rev. Lett.}\ }\textbf {\bibinfo
  {volume} {82}},\ \bibinfo {pages} {3956} (\bibinfo {year} {1999})},\ \Eprint
  {http://arxiv.org/abs/hep-ph/9811473} {arXiv:hep-ph/9811473 [hep-ph]}
  \BibitemShut {NoStop}%
\bibitem [{\citenamefont {Macher}\ and\ \citenamefont
  {Schaffner-Bielich}(2005)}]{Macher:2004vw}%
  \BibitemOpen
  \bibfield  {author} {\bibinfo {author} {\bibfnamefont {J.}~\bibnamefont
  {Macher}}\ and\ \bibinfo {author} {\bibfnamefont {J.}~\bibnamefont
  {Schaffner-Bielich}},\ }\href {\doibase 10.1088/0143-0807/26/3/003}
  {\bibfield  {journal} {\bibinfo  {journal} {Eur. J. Phys.}\ }\textbf
  {\bibinfo {volume} {26}},\ \bibinfo {pages} {341} (\bibinfo {year} {2005})},\
  \Eprint {http://arxiv.org/abs/astro-ph/0411295} {arXiv:astro-ph/0411295
  [astro-ph]} \BibitemShut {NoStop}%
\bibitem [{\citenamefont {Masuda}\ \emph {et~al.}(2013)\citenamefont {Masuda},
  \citenamefont {Hatsuda},\ and\ \citenamefont {Takatsuka}}]{Masuda:2012kf}%
  \BibitemOpen
  \bibfield  {author} {\bibinfo {author} {\bibfnamefont {K.}~\bibnamefont
  {Masuda}}, \bibinfo {author} {\bibfnamefont {T.}~\bibnamefont {Hatsuda}}, \
  and\ \bibinfo {author} {\bibfnamefont {T.}~\bibnamefont {Takatsuka}},\ }\href
  {\doibase 10.1088/0004-637X/764/1/12} {\bibfield  {journal} {\bibinfo
  {journal} {Astrophys. J.}\ }\textbf {\bibinfo {volume} {764}},\ \bibinfo
  {pages} {12} (\bibinfo {year} {2013})},\ \Eprint
  {http://arxiv.org/abs/1205.3621} {arXiv:1205.3621 [nucl-th]} \BibitemShut
  {NoStop}%
\bibitem [{\citenamefont {Heiselberg}\ \emph {et~al.}(1993)\citenamefont
  {Heiselberg}, \citenamefont {Pethick},\ and\ \citenamefont
  {Staubo}}]{Heiselberg:1992dx}%
  \BibitemOpen
  \bibfield  {author} {\bibinfo {author} {\bibfnamefont {H.}~\bibnamefont
  {Heiselberg}}, \bibinfo {author} {\bibfnamefont {C.~J.}\ \bibnamefont
  {Pethick}}, \ and\ \bibinfo {author} {\bibfnamefont {E.~F.}\ \bibnamefont
  {Staubo}},\ }\href {\doibase 10.1103/PhysRevLett.70.1355} {\bibfield
  {journal} {\bibinfo  {journal} {Phys. Rev. Lett.}\ }\textbf {\bibinfo
  {volume} {70}},\ \bibinfo {pages} {1355} (\bibinfo {year}
  {1993})}\BibitemShut {NoStop}%
\bibitem [{\citenamefont {Endo}\ \emph {et~al.}(2006)\citenamefont {Endo},
  \citenamefont {Maruyama}, \citenamefont {Chiba},\ and\ \citenamefont
  {Tatsumi}}]{Endo:2005zt}%
  \BibitemOpen
  \bibfield  {author} {\bibinfo {author} {\bibfnamefont {T.}~\bibnamefont
  {Endo}}, \bibinfo {author} {\bibfnamefont {T.}~\bibnamefont {Maruyama}},
  \bibinfo {author} {\bibfnamefont {S.}~\bibnamefont {Chiba}}, \ and\ \bibinfo
  {author} {\bibfnamefont {T.}~\bibnamefont {Tatsumi}},\ }\href {\doibase
  10.1143/PTP.115.337} {\bibfield  {journal} {\bibinfo  {journal} {Prog. Theor.
  Phys.}\ }\textbf {\bibinfo {volume} {115}},\ \bibinfo {pages} {337} (\bibinfo
  {year} {2006})},\ \Eprint {http://arxiv.org/abs/hep-ph/0510279}
  {arXiv:hep-ph/0510279 [hep-ph]} \BibitemShut {NoStop}%
\bibitem [{\citenamefont {Lugones}\ \emph {et~al.}(2013)\citenamefont
  {Lugones}, \citenamefont {Grunfeld},\ and\ \citenamefont
  {Al~Ajmi}}]{Lugones:2013ema}%
  \BibitemOpen
  \bibfield  {author} {\bibinfo {author} {\bibfnamefont {G.}~\bibnamefont
  {Lugones}}, \bibinfo {author} {\bibfnamefont {A.~G.}\ \bibnamefont
  {Grunfeld}}, \ and\ \bibinfo {author} {\bibfnamefont {M.}~\bibnamefont
  {Al~Ajmi}},\ }\href {\doibase 10.1103/PhysRevC.88.045803} {\bibfield
  {journal} {\bibinfo  {journal} {Phys. Rev.}\ }\textbf {\bibinfo {volume}
  {C88}},\ \bibinfo {pages} {045803} (\bibinfo {year} {2013})},\ \Eprint
  {http://arxiv.org/abs/1308.1452} {arXiv:1308.1452 [hep-ph]} \BibitemShut
  {NoStop}%
\bibitem [{\citenamefont {Contrera}\ \emph {et~al.}(2014)\citenamefont
  {Contrera}, \citenamefont {Spinella}, \citenamefont {Orsaria},\ and\
  \citenamefont {Weber}}]{Contrera:2014eza}%
  \BibitemOpen
  \bibfield  {author} {\bibinfo {author} {\bibfnamefont {G.~A.}\ \bibnamefont
  {Contrera}}, \bibinfo {author} {\bibfnamefont {W.}~\bibnamefont {Spinella}},
  \bibinfo {author} {\bibfnamefont {M.}~\bibnamefont {Orsaria}}, \ and\
  \bibinfo {author} {\bibfnamefont {F.}~\bibnamefont {Weber}},\ }in\ \href@noop
  {} {\emph {\bibinfo {booktitle} {{6th International Workshop on Astronomy and
  Relativistic Astrophysics Rio de Janeiro, Rio de Janeiro, Brazil, September
  29-October 3, 2013}}}}\ (\bibinfo {year} {2014})\ \Eprint
  {http://arxiv.org/abs/1403.7415} {arXiv:1403.7415 [hep-ph]} \BibitemShut
  {NoStop}%
\bibitem [{\citenamefont {Benic}\ \emph {et~al.}(2015)\citenamefont {Benic},
  \citenamefont {Blaschke}, \citenamefont {Alvarez-Castillo}, \citenamefont
  {Fischer},\ and\ \citenamefont {Typel}}]{Benic:2014jia}%
  \BibitemOpen
  \bibfield  {author} {\bibinfo {author} {\bibfnamefont {S.}~\bibnamefont
  {Benic}}, \bibinfo {author} {\bibfnamefont {D.}~\bibnamefont {Blaschke}},
  \bibinfo {author} {\bibfnamefont {D.~E.}\ \bibnamefont {Alvarez-Castillo}},
  \bibinfo {author} {\bibfnamefont {T.}~\bibnamefont {Fischer}}, \ and\
  \bibinfo {author} {\bibfnamefont {S.}~\bibnamefont {Typel}},\ }\href
  {\doibase 10.1051/0004-6361/201425318} {\bibfield  {journal} {\bibinfo
  {journal} {Astron. Astrophys.}\ }\textbf {\bibinfo {volume} {577}},\ \bibinfo
  {pages} {A40} (\bibinfo {year} {2015})},\ \Eprint
  {http://arxiv.org/abs/1411.2856} {arXiv:1411.2856 [astro-ph.HE]} \BibitemShut
  {NoStop}%
\bibitem [{\citenamefont {Li}\ \emph {et~al.}(2018)\citenamefont {Li},
  \citenamefont {Zhang}, \citenamefont {Yan}, \citenamefont {Huang},\ and\
  \citenamefont {Zong}}]{Li:2018ltg}%
  \BibitemOpen
  \bibfield  {author} {\bibinfo {author} {\bibfnamefont {C.-M.}\ \bibnamefont
  {Li}}, \bibinfo {author} {\bibfnamefont {J.-L.}\ \bibnamefont {Zhang}},
  \bibinfo {author} {\bibfnamefont {Y.}~\bibnamefont {Yan}}, \bibinfo {author}
  {\bibfnamefont {Y.-F.}\ \bibnamefont {Huang}}, \ and\ \bibinfo {author}
  {\bibfnamefont {H.-S.}\ \bibnamefont {Zong}},\ }\href {\doibase
  10.1103/PhysRevD.97.103013} {\bibfield  {journal} {\bibinfo  {journal} {Phys.
  Rev.}\ }\textbf {\bibinfo {volume} {D97}},\ \bibinfo {pages} {103013}
  (\bibinfo {year} {2018})},\ \Eprint {http://arxiv.org/abs/1804.10785}
  {arXiv:1804.10785 [nucl-th]} \BibitemShut {NoStop}%
\bibitem [{\citenamefont {Abgaryan}\ \emph {et~al.}(2018)\citenamefont
  {Abgaryan}, \citenamefont {Alvarez-Castillo}, \citenamefont {Ayriyan},
  \citenamefont {Blaschke},\ and\ \citenamefont
  {Grigorian}}]{Abgaryan:2018gqp}%
  \BibitemOpen
  \bibfield  {author} {\bibinfo {author} {\bibfnamefont {V.}~\bibnamefont
  {Abgaryan}}, \bibinfo {author} {\bibfnamefont {D.}~\bibnamefont
  {Alvarez-Castillo}}, \bibinfo {author} {\bibfnamefont {A.}~\bibnamefont
  {Ayriyan}}, \bibinfo {author} {\bibfnamefont {D.}~\bibnamefont {Blaschke}}, \
  and\ \bibinfo {author} {\bibfnamefont {H.}~\bibnamefont {Grigorian}},\ }\href
  {\doibase 10.3390/universe4090094} {\bibfield  {journal} {\bibinfo  {journal}
  {Universe}\ }\textbf {\bibinfo {volume} {4}},\ \bibinfo {pages} {94}
  (\bibinfo {year} {2018})},\ \Eprint {http://arxiv.org/abs/1807.08034}
  {arXiv:1807.08034 [astro-ph.HE]} \BibitemShut {NoStop}%
\bibitem [{\citenamefont {Kaltenborn}\ \emph {et~al.}(2017)\citenamefont
  {Kaltenborn}, \citenamefont {Bastian},\ and\ \citenamefont
  {Blaschke}}]{Kaltenborn:2017hus}%
  \BibitemOpen
  \bibfield  {author} {\bibinfo {author} {\bibfnamefont {M.~A.~R.}\
  \bibnamefont {Kaltenborn}}, \bibinfo {author} {\bibfnamefont {N.-U.~F.}\
  \bibnamefont {Bastian}}, \ and\ \bibinfo {author} {\bibfnamefont {D.~B.}\
  \bibnamefont {Blaschke}},\ }\href {\doibase 10.1103/PhysRevD.96.056024}
  {\bibfield  {journal} {\bibinfo  {journal} {Phys. Rev.}\ }\textbf {\bibinfo
  {volume} {D96}},\ \bibinfo {pages} {056024} (\bibinfo {year} {2017})},\
  \Eprint {http://arxiv.org/abs/1701.04400} {arXiv:1701.04400 [astro-ph.HE]}
  \BibitemShut {NoStop}%
\bibitem [{\citenamefont {Hempel}\ \emph {et~al.}(2013)\citenamefont {Hempel},
  \citenamefont {Dexheimer}, \citenamefont {Schramm},\ and\ \citenamefont
  {Iosilevskiy}}]{Hempel:2013tfa}%
  \BibitemOpen
  \bibfield  {author} {\bibinfo {author} {\bibfnamefont {M.}~\bibnamefont
  {Hempel}}, \bibinfo {author} {\bibfnamefont {V.}~\bibnamefont {Dexheimer}},
  \bibinfo {author} {\bibfnamefont {S.}~\bibnamefont {Schramm}}, \ and\
  \bibinfo {author} {\bibfnamefont {I.}~\bibnamefont {Iosilevskiy}},\ }\href
  {\doibase 10.1103/PhysRevC.88.014906} {\bibfield  {journal} {\bibinfo
  {journal} {Phys. Rev.}\ }\textbf {\bibinfo {volume} {C88}},\ \bibinfo {pages}
  {014906} (\bibinfo {year} {2013})},\ \Eprint {http://arxiv.org/abs/1302.2835}
  {arXiv:1302.2835 [nucl-th]} \BibitemShut {NoStop}%
\bibitem [{\citenamefont {Dexheimer}\ \emph {et~al.}(2018)\citenamefont
  {Dexheimer}, \citenamefont {Soethe}, \citenamefont {Roark}, \citenamefont
  {Gomes}, \citenamefont {Kepler},\ and\ \citenamefont
  {Schramm}}]{Dexheimer:2019pay}%
  \BibitemOpen
  \bibfield  {author} {\bibinfo {author} {\bibfnamefont {V.}~\bibnamefont
  {Dexheimer}}, \bibinfo {author} {\bibfnamefont {L.~T.~T.}\ \bibnamefont
  {Soethe}}, \bibinfo {author} {\bibfnamefont {J.}~\bibnamefont {Roark}},
  \bibinfo {author} {\bibfnamefont {R.~O.}\ \bibnamefont {Gomes}}, \bibinfo
  {author} {\bibfnamefont {S.~O.}\ \bibnamefont {Kepler}}, \ and\ \bibinfo
  {author} {\bibfnamefont {S.}~\bibnamefont {Schramm}},\ }\href {\doibase
  10.1142/S0218301318300084} {\bibfield  {journal} {\bibinfo  {journal} {Int.
  J. Mod. Phys.}\ }\textbf {\bibinfo {volume} {E27}},\ \bibinfo {pages}
  {1830008} (\bibinfo {year} {2018})},\ \Eprint
  {http://arxiv.org/abs/1901.03252} {arXiv:1901.03252 [astro-ph.HE]}
  \BibitemShut {NoStop}%
\bibitem [{\citenamefont {Cloët}\ \emph
  {et~al.}(2005{\natexlab{a}})\citenamefont {Cloët}, \citenamefont {Bentz},\
  and\ \citenamefont {Thomas}}]{Cloet:2005pp}%
  \BibitemOpen
  \bibfield  {author} {\bibinfo {author} {\bibfnamefont {I.~C.}\ \bibnamefont
  {Cloët}}, \bibinfo {author} {\bibfnamefont {W.}~\bibnamefont {Bentz}}, \
  and\ \bibinfo {author} {\bibfnamefont {A.~W.}\ \bibnamefont {Thomas}},\
  }\href {\doibase 10.1016/j.physletb.2005.06.065} {\bibfield  {journal}
  {\bibinfo  {journal} {Phys. Lett.}\ }\textbf {\bibinfo {volume} {B621}},\
  \bibinfo {pages} {246} (\bibinfo {year} {2005}{\natexlab{a}})},\ \Eprint
  {http://arxiv.org/abs/hep-ph/0504229} {arXiv:hep-ph/0504229 [hep-ph]}
  \BibitemShut {NoStop}%
\bibitem [{\citenamefont {Cloët}\ \emph {et~al.}(2008)\citenamefont {Cloët},
  \citenamefont {Bentz},\ and\ \citenamefont {Thomas}}]{Cloet:2007em}%
  \BibitemOpen
  \bibfield  {author} {\bibinfo {author} {\bibfnamefont {I.~C.}\ \bibnamefont
  {Cloët}}, \bibinfo {author} {\bibfnamefont {W.}~\bibnamefont {Bentz}}, \
  and\ \bibinfo {author} {\bibfnamefont {A.~W.}\ \bibnamefont {Thomas}},\
  }\href {\doibase 10.1016/j.physletb.2007.09.071} {\bibfield  {journal}
  {\bibinfo  {journal} {Phys. Lett.}\ }\textbf {\bibinfo {volume} {B659}},\
  \bibinfo {pages} {214} (\bibinfo {year} {2008})},\ \Eprint
  {http://arxiv.org/abs/0708.3246} {arXiv:0708.3246 [hep-ph]} \BibitemShut
  {NoStop}%
\bibitem [{\citenamefont {Cloët}\ \emph {et~al.}(2014)\citenamefont {Cloët},
  \citenamefont {Bentz},\ and\ \citenamefont {Thomas}}]{Cloet:2014rja}%
  \BibitemOpen
  \bibfield  {author} {\bibinfo {author} {\bibfnamefont {I.~C.}\ \bibnamefont
  {Cloët}}, \bibinfo {author} {\bibfnamefont {W.}~\bibnamefont {Bentz}}, \
  and\ \bibinfo {author} {\bibfnamefont {A.~W.}\ \bibnamefont {Thomas}},\
  }\href {\doibase 10.1103/PhysRevC.90.045202} {\bibfield  {journal} {\bibinfo
  {journal} {Phys. Rev.}\ }\textbf {\bibinfo {volume} {C90}},\ \bibinfo {pages}
  {045202} (\bibinfo {year} {2014})},\ \Eprint {http://arxiv.org/abs/1405.5542}
  {arXiv:1405.5542 [nucl-th]} \BibitemShut {NoStop}%
\bibitem [{\citenamefont {Hutauruk}\ \emph {et~al.}(2016)\citenamefont
  {Hutauruk}, \citenamefont {Cloët},\ and\ \citenamefont
  {Thomas}}]{Hutauruk:2016sug}%
  \BibitemOpen
  \bibfield  {author} {\bibinfo {author} {\bibfnamefont {P.~T.~P.}\
  \bibnamefont {Hutauruk}}, \bibinfo {author} {\bibfnamefont {I.~C.}\
  \bibnamefont {Cloët}}, \ and\ \bibinfo {author} {\bibfnamefont {A.~W.}\
  \bibnamefont {Thomas}},\ }\href {\doibase 10.1103/PhysRevC.94.035201}
  {\bibfield  {journal} {\bibinfo  {journal} {Phys. Rev.}\ }\textbf {\bibinfo
  {volume} {C94}},\ \bibinfo {pages} {035201} (\bibinfo {year} {2016})},\
  \Eprint {http://arxiv.org/abs/1604.02853} {arXiv:1604.02853 [nucl-th]}
  \BibitemShut {NoStop}%
\bibitem [{\citenamefont {Ninomiya}\ \emph {et~al.}(2017)\citenamefont
  {Ninomiya}, \citenamefont {Bentz},\ and\ \citenamefont
  {Cloët}}]{Ninomiya:2017ggn}%
  \BibitemOpen
  \bibfield  {author} {\bibinfo {author} {\bibfnamefont {Y.}~\bibnamefont
  {Ninomiya}}, \bibinfo {author} {\bibfnamefont {W.}~\bibnamefont {Bentz}}, \
  and\ \bibinfo {author} {\bibfnamefont {I.~C.}\ \bibnamefont {Cloët}},\
  }\href {\doibase 10.1103/PhysRevC.96.045206} {\bibfield  {journal} {\bibinfo
  {journal} {Phys. Rev.}\ }\textbf {\bibinfo {volume} {C96}},\ \bibinfo {pages}
  {045206} (\bibinfo {year} {2017})},\ \Eprint
  {http://arxiv.org/abs/1707.03787} {arXiv:1707.03787 [nucl-th]} \BibitemShut
  {NoStop}%
\bibitem [{\citenamefont {Cloët}\ \emph
  {et~al.}(2005{\natexlab{b}})\citenamefont {Cloët}, \citenamefont {Bentz},\
  and\ \citenamefont {Thomas}}]{Cloet:2005rt}%
  \BibitemOpen
  \bibfield  {author} {\bibinfo {author} {\bibfnamefont {I.~C.}\ \bibnamefont
  {Cloët}}, \bibinfo {author} {\bibfnamefont {W.}~\bibnamefont {Bentz}}, \
  and\ \bibinfo {author} {\bibfnamefont {A.~W.}\ \bibnamefont {Thomas}},\
  }\href {\doibase 10.1103/PhysRevLett.95.052302} {\bibfield  {journal}
  {\bibinfo  {journal} {Phys. Rev. Lett.}\ }\textbf {\bibinfo {volume} {95}},\
  \bibinfo {pages} {052302} (\bibinfo {year} {2005}{\natexlab{b}})},\ \Eprint
  {http://arxiv.org/abs/nucl-th/0504019} {arXiv:nucl-th/0504019 [nucl-th]}
  \BibitemShut {NoStop}%
\bibitem [{\citenamefont {Cloët}\ \emph {et~al.}(2006)\citenamefont {Cloët},
  \citenamefont {Bentz},\ and\ \citenamefont {Thomas}}]{Cloet:2006bq}%
  \BibitemOpen
  \bibfield  {author} {\bibinfo {author} {\bibfnamefont {I.~C.}\ \bibnamefont
  {Cloët}}, \bibinfo {author} {\bibfnamefont {W.}~\bibnamefont {Bentz}}, \
  and\ \bibinfo {author} {\bibfnamefont {A.~W.}\ \bibnamefont {Thomas}},\
  }\href {\doibase 10.1016/j.physletb.2006.08.076} {\bibfield  {journal}
  {\bibinfo  {journal} {Phys. Lett.}\ }\textbf {\bibinfo {volume} {B642}},\
  \bibinfo {pages} {210} (\bibinfo {year} {2006})},\ \Eprint
  {http://arxiv.org/abs/nucl-th/0605061} {arXiv:nucl-th/0605061 [nucl-th]}
  \BibitemShut {NoStop}%
\bibitem [{\citenamefont {Cloët}\ \emph {et~al.}(2016)\citenamefont {Cloët},
  \citenamefont {Bentz},\ and\ \citenamefont {Thomas}}]{Cloet:2015tha}%
  \BibitemOpen
  \bibfield  {author} {\bibinfo {author} {\bibfnamefont {I.~C.}\ \bibnamefont
  {Cloët}}, \bibinfo {author} {\bibfnamefont {W.}~\bibnamefont {Bentz}}, \
  and\ \bibinfo {author} {\bibfnamefont {A.~W.}\ \bibnamefont {Thomas}},\
  }\href {\doibase 10.1103/PhysRevLett.116.032701} {\bibfield  {journal}
  {\bibinfo  {journal} {Phys. Rev. Lett.}\ }\textbf {\bibinfo {volume} {116}},\
  \bibinfo {pages} {032701} (\bibinfo {year} {2016})},\ \Eprint
  {http://arxiv.org/abs/1506.05875} {arXiv:1506.05875 [nucl-th]} \BibitemShut
  {NoStop}%
\bibitem [{\citenamefont {Hellstern}\ \emph {et~al.}(1997)\citenamefont
  {Hellstern}, \citenamefont {Alkofer},\ and\ \citenamefont
  {Reinhardt}}]{Hellstern:1997nv}%
  \BibitemOpen
  \bibfield  {author} {\bibinfo {author} {\bibfnamefont {G.}~\bibnamefont
  {Hellstern}}, \bibinfo {author} {\bibfnamefont {R.}~\bibnamefont {Alkofer}},
  \ and\ \bibinfo {author} {\bibfnamefont {H.}~\bibnamefont {Reinhardt}},\
  }\href {\doibase 10.1016/S0375-9474(97)00412-0} {\bibfield  {journal}
  {\bibinfo  {journal} {Nucl. Phys.}\ }\textbf {\bibinfo {volume} {A625}},\
  \bibinfo {pages} {697} (\bibinfo {year} {1997})},\ \Eprint
  {http://arxiv.org/abs/hep-ph/9706551} {arXiv:hep-ph/9706551 [hep-ph]}
  \BibitemShut {NoStop}%
\bibitem [{\citenamefont {Bentz}\ and\ \citenamefont
  {Thomas}(2001)}]{Bentz:2001vc}%
  \BibitemOpen
  \bibfield  {author} {\bibinfo {author} {\bibfnamefont {W.}~\bibnamefont
  {Bentz}}\ and\ \bibinfo {author} {\bibfnamefont {A.~W.}\ \bibnamefont
  {Thomas}},\ }\href {\doibase 10.1016/S0375-9474(01)01119-8} {\bibfield
  {journal} {\bibinfo  {journal} {Nucl. Phys.}\ }\textbf {\bibinfo {volume}
  {A696}},\ \bibinfo {pages} {138} (\bibinfo {year} {2001})},\ \Eprint
  {http://arxiv.org/abs/nucl-th/0105022} {arXiv:nucl-th/0105022 [nucl-th]}
  \BibitemShut {NoStop}%
\bibitem [{\citenamefont {Alford}\ \emph {et~al.}(2013)\citenamefont {Alford},
  \citenamefont {Han},\ and\ \citenamefont {Prakash}}]{Alford:2013aca}%
  \BibitemOpen
  \bibfield  {author} {\bibinfo {author} {\bibfnamefont {M.~G.}\ \bibnamefont
  {Alford}}, \bibinfo {author} {\bibfnamefont {S.}~\bibnamefont {Han}}, \ and\
  \bibinfo {author} {\bibfnamefont {M.}~\bibnamefont {Prakash}},\ }\href
  {\doibase 10.1103/PhysRevD.88.083013} {\bibfield  {journal} {\bibinfo
  {journal} {Phys. Rev.}\ }\textbf {\bibinfo {volume} {D88}},\ \bibinfo {pages}
  {083013} (\bibinfo {year} {2013})},\ \Eprint {http://arxiv.org/abs/1302.4732}
  {arXiv:1302.4732 [astro-ph.SR]} \BibitemShut {NoStop}%
\bibitem [{\citenamefont {Ranea-Sandoval}\ \emph {et~al.}(2016)\citenamefont
  {Ranea-Sandoval}, \citenamefont {Han}, \citenamefont {Orsaria}, \citenamefont
  {Contrera}, \citenamefont {Weber},\ and\ \citenamefont
  {Alford}}]{Ranea-Sandoval:2015ldr}%
  \BibitemOpen
  \bibfield  {author} {\bibinfo {author} {\bibfnamefont {I.~F.}\ \bibnamefont
  {Ranea-Sandoval}}, \bibinfo {author} {\bibfnamefont {S.}~\bibnamefont {Han}},
  \bibinfo {author} {\bibfnamefont {M.~G.}\ \bibnamefont {Orsaria}}, \bibinfo
  {author} {\bibfnamefont {G.~A.}\ \bibnamefont {Contrera}}, \bibinfo {author}
  {\bibfnamefont {F.}~\bibnamefont {Weber}}, \ and\ \bibinfo {author}
  {\bibfnamefont {M.~G.}\ \bibnamefont {Alford}},\ }\href {\doibase
  10.1103/PhysRevC.93.045812} {\bibfield  {journal} {\bibinfo  {journal} {Phys.
  Rev.}\ }\textbf {\bibinfo {volume} {C93}},\ \bibinfo {pages} {045812}
  (\bibinfo {year} {2016})},\ \Eprint {http://arxiv.org/abs/1512.09183}
  {arXiv:1512.09183 [nucl-th]} \BibitemShut {NoStop}%
\bibitem [{\citenamefont {Whittenbury}\ \emph {et~al.}(2016)\citenamefont
  {Whittenbury}, \citenamefont {Matevosyan},\ and\ \citenamefont
  {Thomas}}]{Whittenbury:2015ziz}%
  \BibitemOpen
  \bibfield  {author} {\bibinfo {author} {\bibfnamefont {D.~L.}\ \bibnamefont
  {Whittenbury}}, \bibinfo {author} {\bibfnamefont {H.~H.}\ \bibnamefont
  {Matevosyan}}, \ and\ \bibinfo {author} {\bibfnamefont {A.~W.}\ \bibnamefont
  {Thomas}},\ }\href {\doibase 10.1103/PhysRevC.93.035807} {\bibfield
  {journal} {\bibinfo  {journal} {Phys. Rev.}\ }\textbf {\bibinfo {volume}
  {C93}},\ \bibinfo {pages} {035807} (\bibinfo {year} {2016})},\ \Eprint
  {http://arxiv.org/abs/1511.08561} {arXiv:1511.08561 [nucl-th]} \BibitemShut
  {NoStop}%
\bibitem [{\citenamefont {Hell}\ \emph {et~al.}(2013)\citenamefont {Hell},
  \citenamefont {Kashiwa},\ and\ \citenamefont {Weise}}]{Hell:2012da}%
  \BibitemOpen
  \bibfield  {author} {\bibinfo {author} {\bibfnamefont {T.}~\bibnamefont
  {Hell}}, \bibinfo {author} {\bibfnamefont {K.}~\bibnamefont {Kashiwa}}, \
  and\ \bibinfo {author} {\bibfnamefont {W.}~\bibnamefont {Weise}},\ }\href
  {\doibase 10.4236/jmp.2013.45093} {\bibfield  {journal} {\bibinfo  {journal}
  {J. Mod. Phys.}\ }\textbf {\bibinfo {volume} {4}},\ \bibinfo {pages} {644}
  (\bibinfo {year} {2013})},\ \Eprint {http://arxiv.org/abs/1212.4017}
  {arXiv:1212.4017 [hep-ph]} \BibitemShut {NoStop}%
\bibitem [{\citenamefont {Cloët}\ \emph {et~al.}(2009)\citenamefont {Cloët},
  \citenamefont {Bentz},\ and\ \citenamefont {Thomas}}]{Cloet:2009qs}%
  \BibitemOpen
  \bibfield  {author} {\bibinfo {author} {\bibfnamefont {I.~C.}\ \bibnamefont
  {Cloët}}, \bibinfo {author} {\bibfnamefont {W.}~\bibnamefont {Bentz}}, \
  and\ \bibinfo {author} {\bibfnamefont {A.~W.}\ \bibnamefont {Thomas}},\
  }\href {\doibase 10.1103/PhysRevLett.102.252301} {\bibfield  {journal}
  {\bibinfo  {journal} {Phys. Rev. Lett.}\ }\textbf {\bibinfo {volume} {102}},\
  \bibinfo {pages} {252301} (\bibinfo {year} {2009})},\ \Eprint
  {http://arxiv.org/abs/0901.3559} {arXiv:0901.3559 [nucl-th]} \BibitemShut
  {NoStop}%
\bibitem [{\citenamefont {Cloët}\ \emph {et~al.}(2012)\citenamefont {Cloët},
  \citenamefont {Bentz},\ and\ \citenamefont {Thomas}}]{Cloet:2012td}%
  \BibitemOpen
  \bibfield  {author} {\bibinfo {author} {\bibfnamefont {I.~C.}\ \bibnamefont
  {Cloët}}, \bibinfo {author} {\bibfnamefont {W.}~\bibnamefont {Bentz}}, \
  and\ \bibinfo {author} {\bibfnamefont {A.~W.}\ \bibnamefont {Thomas}},\
  }\href {\doibase 10.1103/PhysRevLett.109.182301} {\bibfield  {journal}
  {\bibinfo  {journal} {Phys. Rev. Lett.}\ }\textbf {\bibinfo {volume} {109}},\
  \bibinfo {pages} {182301} (\bibinfo {year} {2012})},\ \Eprint
  {http://arxiv.org/abs/1202.6401} {arXiv:1202.6401 [nucl-th]} \BibitemShut
  {NoStop}%
\bibitem [{\citenamefont {Ruester}\ \emph {et~al.}(2005)\citenamefont
  {Ruester}, \citenamefont {Werth}, \citenamefont {Buballa}, \citenamefont
  {Shovkovy},\ and\ \citenamefont {Rischke}}]{Ruester:2005jc}%
  \BibitemOpen
  \bibfield  {author} {\bibinfo {author} {\bibfnamefont {S.~B.}\ \bibnamefont
  {Ruester}}, \bibinfo {author} {\bibfnamefont {V.}~\bibnamefont {Werth}},
  \bibinfo {author} {\bibfnamefont {M.}~\bibnamefont {Buballa}}, \bibinfo
  {author} {\bibfnamefont {I.~A.}\ \bibnamefont {Shovkovy}}, \ and\ \bibinfo
  {author} {\bibfnamefont {D.~H.}\ \bibnamefont {Rischke}},\ }\href {\doibase
  10.1103/PhysRevD.72.034004} {\bibfield  {journal} {\bibinfo  {journal} {Phys.
  Rev.}\ }\textbf {\bibinfo {volume} {D72}},\ \bibinfo {pages} {034004}
  (\bibinfo {year} {2005})},\ \Eprint {http://arxiv.org/abs/hep-ph/0503184}
  {arXiv:hep-ph/0503184 [hep-ph]} \BibitemShut {NoStop}%
\bibitem [{\citenamefont {Abuki}\ and\ \citenamefont
  {Kunihiro}(2006)}]{Abuki:2005ms}%
  \BibitemOpen
  \bibfield  {author} {\bibinfo {author} {\bibfnamefont {H.}~\bibnamefont
  {Abuki}}\ and\ \bibinfo {author} {\bibfnamefont {T.}~\bibnamefont
  {Kunihiro}},\ }\href {\doibase 10.1016/j.nuclphysa.2005.12.019} {\bibfield
  {journal} {\bibinfo  {journal} {Nucl. Phys.}\ }\textbf {\bibinfo {volume}
  {A768}},\ \bibinfo {pages} {118} (\bibinfo {year} {2006})},\ \Eprint
  {http://arxiv.org/abs/hep-ph/0509172} {arXiv:hep-ph/0509172 [hep-ph]}
  \BibitemShut {NoStop}%
\bibitem [{\citenamefont {Steiner}\ \emph {et~al.}(2002)\citenamefont
  {Steiner}, \citenamefont {Reddy},\ and\ \citenamefont
  {Prakash}}]{Steiner:2002gx}%
  \BibitemOpen
  \bibfield  {author} {\bibinfo {author} {\bibfnamefont {A.~W.}\ \bibnamefont
  {Steiner}}, \bibinfo {author} {\bibfnamefont {S.}~\bibnamefont {Reddy}}, \
  and\ \bibinfo {author} {\bibfnamefont {M.}~\bibnamefont {Prakash}},\ }\href
  {\doibase 10.1103/PhysRevD.66.094007} {\bibfield  {journal} {\bibinfo
  {journal} {Phys. Rev.}\ }\textbf {\bibinfo {volume} {D66}},\ \bibinfo {pages}
  {094007} (\bibinfo {year} {2002})},\ \Eprint
  {http://arxiv.org/abs/hep-ph/0205201} {arXiv:hep-ph/0205201 [hep-ph]}
  \BibitemShut {NoStop}%
\bibitem [{\citenamefont {Carrillo-Serrano}\ \emph {et~al.}(2016)\citenamefont
  {Carrillo-Serrano}, \citenamefont {Bentz}, \citenamefont {Cloët},\ and\
  \citenamefont {Thomas}}]{Carrillo-Serrano:2016igi}%
  \BibitemOpen
  \bibfield  {author} {\bibinfo {author} {\bibfnamefont {M.~E.}\ \bibnamefont
  {Carrillo-Serrano}}, \bibinfo {author} {\bibfnamefont {W.}~\bibnamefont
  {Bentz}}, \bibinfo {author} {\bibfnamefont {I.~C.}\ \bibnamefont {Cloët}}, \
  and\ \bibinfo {author} {\bibfnamefont {A.~W.}\ \bibnamefont {Thomas}},\
  }\href {\doibase 10.1016/j.physletb.2016.05.065} {\bibfield  {journal}
  {\bibinfo  {journal} {Phys. Lett.}\ }\textbf {\bibinfo {volume} {B759}},\
  \bibinfo {pages} {178} (\bibinfo {year} {2016})},\ \Eprint
  {http://arxiv.org/abs/1603.02741} {arXiv:1603.02741 [nucl-th]} \BibitemShut
  {NoStop}%
\bibitem [{\citenamefont {Endo}(2011)}]{Endo:2011em}%
  \BibitemOpen
  \bibfield  {author} {\bibinfo {author} {\bibfnamefont {T.}~\bibnamefont
  {Endo}},\ }\href {\doibase 10.1103/PhysRevC.83.068801} {\bibfield  {journal}
  {\bibinfo  {journal} {Phys. Rev.}\ }\textbf {\bibinfo {volume} {C83}},\
  \bibinfo {pages} {068801} (\bibinfo {year} {2011})},\ \Eprint
  {http://arxiv.org/abs/1105.2445} {arXiv:1105.2445 [astro-ph.SR]} \BibitemShut
  {NoStop}%
\bibitem [{\citenamefont {Neumann}\ \emph {et~al.}(2003)\citenamefont
  {Neumann}, \citenamefont {Buballa},\ and\ \citenamefont
  {Oertel}}]{Neumann:2002jm}%
  \BibitemOpen
  \bibfield  {author} {\bibinfo {author} {\bibfnamefont {F.}~\bibnamefont
  {Neumann}}, \bibinfo {author} {\bibfnamefont {M.}~\bibnamefont {Buballa}}, \
  and\ \bibinfo {author} {\bibfnamefont {M.}~\bibnamefont {Oertel}},\ }\href
  {\doibase 10.1016/S0375-9474(02)01371-4} {\bibfield  {journal} {\bibinfo
  {journal} {Nucl. Phys.}\ }\textbf {\bibinfo {volume} {A714}},\ \bibinfo
  {pages} {481} (\bibinfo {year} {2003})},\ \Eprint
  {http://arxiv.org/abs/hep-ph/0210078} {arXiv:hep-ph/0210078 [hep-ph]}
  \BibitemShut {NoStop}%
\bibitem [{\citenamefont {Voskresensky}\ \emph {et~al.}(2003)\citenamefont
  {Voskresensky}, \citenamefont {Yasuhira},\ and\ \citenamefont
  {Tatsumi}}]{Voskresensky:2002hu}%
  \BibitemOpen
  \bibfield  {author} {\bibinfo {author} {\bibfnamefont {D.~N.}\ \bibnamefont
  {Voskresensky}}, \bibinfo {author} {\bibfnamefont {M.}~\bibnamefont
  {Yasuhira}}, \ and\ \bibinfo {author} {\bibfnamefont {T.}~\bibnamefont
  {Tatsumi}},\ }\href {\doibase 10.1016/S0375-9474(03)01313-7} {\bibfield
  {journal} {\bibinfo  {journal} {Nucl. Phys.}\ }\textbf {\bibinfo {volume}
  {A723}},\ \bibinfo {pages} {291} (\bibinfo {year} {2003})},\ \Eprint
  {http://arxiv.org/abs/nucl-th/0208067} {arXiv:nucl-th/0208067 [nucl-th]}
  \BibitemShut {NoStop}%
\bibitem [{\citenamefont {Pinto}\ \emph {et~al.}(2012)\citenamefont {Pinto},
  \citenamefont {Koch},\ and\ \citenamefont {Randrup}}]{Pinto:2012aq}%
  \BibitemOpen
  \bibfield  {author} {\bibinfo {author} {\bibfnamefont {M.~B.}\ \bibnamefont
  {Pinto}}, \bibinfo {author} {\bibfnamefont {V.}~\bibnamefont {Koch}}, \ and\
  \bibinfo {author} {\bibfnamefont {J.}~\bibnamefont {Randrup}},\ }\href
  {\doibase 10.1103/PhysRevC.86.025203} {\bibfield  {journal} {\bibinfo
  {journal} {Phys. Rev.}\ }\textbf {\bibinfo {volume} {C86}},\ \bibinfo {pages}
  {025203} (\bibinfo {year} {2012})},\ \Eprint {http://arxiv.org/abs/1207.5186}
  {arXiv:1207.5186 [hep-ph]} \BibitemShut {NoStop}%
\bibitem [{\citenamefont {Fraga}\ \emph {et~al.}(2019)\citenamefont {Fraga},
  \citenamefont {Hippert},\ and\ \citenamefont {Schmitt}}]{Fraga:2018cvr}%
  \BibitemOpen
  \bibfield  {author} {\bibinfo {author} {\bibfnamefont {E.~S.}\ \bibnamefont
  {Fraga}}, \bibinfo {author} {\bibfnamefont {M.}~\bibnamefont {Hippert}}, \
  and\ \bibinfo {author} {\bibfnamefont {A.}~\bibnamefont {Schmitt}},\ }\href
  {\doibase 10.1103/PhysRevD.99.014046} {\bibfield  {journal} {\bibinfo
  {journal} {Phys. Rev.}\ }\textbf {\bibinfo {volume} {D99}},\ \bibinfo {pages}
  {014046} (\bibinfo {year} {2019})},\ \Eprint
  {http://arxiv.org/abs/1810.13226} {arXiv:1810.13226 [hep-ph]} \BibitemShut
  {NoStop}%
\bibitem [{\citenamefont {Dosch}\ and\ \citenamefont
  {Narison}(1998)}]{Dosch:1997wb}%
  \BibitemOpen
  \bibfield  {author} {\bibinfo {author} {\bibfnamefont {H.~G.}\ \bibnamefont
  {Dosch}}\ and\ \bibinfo {author} {\bibfnamefont {S.}~\bibnamefont
  {Narison}},\ }\href {\doibase 10.1016/S0370-2693(97)01370-1} {\bibfield
  {journal} {\bibinfo  {journal} {Phys. Lett.}\ }\textbf {\bibinfo {volume}
  {B417}},\ \bibinfo {pages} {173} (\bibinfo {year} {1998})},\ \Eprint
  {http://arxiv.org/abs/hep-ph/9709215} {arXiv:hep-ph/9709215 [hep-ph]}
  \BibitemShut {NoStop}%
\bibitem [{\citenamefont {Ninomiya}\ \emph {et~al.}(2015)\citenamefont
  {Ninomiya}, \citenamefont {Bentz},\ and\ \citenamefont
  {Cloët}}]{Ninomiya:2014kja}%
  \BibitemOpen
  \bibfield  {author} {\bibinfo {author} {\bibfnamefont {Y.}~\bibnamefont
  {Ninomiya}}, \bibinfo {author} {\bibfnamefont {W.}~\bibnamefont {Bentz}}, \
  and\ \bibinfo {author} {\bibfnamefont {I.~C.}\ \bibnamefont {Cloët}},\
  }\href {\doibase 10.1103/PhysRevC.91.025202} {\bibfield  {journal} {\bibinfo
  {journal} {Phys. Rev.}\ }\textbf {\bibinfo {volume} {C91}},\ \bibinfo {pages}
  {025202} (\bibinfo {year} {2015})},\ \Eprint {http://arxiv.org/abs/1406.7212}
  {arXiv:1406.7212 [nucl-th]} \BibitemShut {NoStop}%
\bibitem [{\citenamefont {Aguilera}\ \emph {et~al.}(2005)\citenamefont
  {Aguilera}, \citenamefont {Blaschke},\ and\ \citenamefont
  {Grigorian}}]{Aguilera:2004ag}%
  \BibitemOpen
  \bibfield  {author} {\bibinfo {author} {\bibfnamefont {D.~N.}\ \bibnamefont
  {Aguilera}}, \bibinfo {author} {\bibfnamefont {D.}~\bibnamefont {Blaschke}},
  \ and\ \bibinfo {author} {\bibfnamefont {H.}~\bibnamefont {Grigorian}},\
  }\href {\doibase 10.1016/j.nuclphysa.2005.04.009} {\bibfield  {journal}
  {\bibinfo  {journal} {Nucl. Phys.}\ }\textbf {\bibinfo {volume} {A757}},\
  \bibinfo {pages} {527} (\bibinfo {year} {2005})},\ \Eprint
  {http://arxiv.org/abs/hep-ph/0412266} {arXiv:hep-ph/0412266 [hep-ph]}
  \BibitemShut {NoStop}%
\bibitem [{\citenamefont {Bentz}\ \emph {et~al.}(2003)\citenamefont {Bentz},
  \citenamefont {Horikawa}, \citenamefont {Ishii},\ and\ \citenamefont
  {Thomas}}]{Bentz:2002um}%
  \BibitemOpen
  \bibfield  {author} {\bibinfo {author} {\bibfnamefont {W.}~\bibnamefont
  {Bentz}}, \bibinfo {author} {\bibfnamefont {T.}~\bibnamefont {Horikawa}},
  \bibinfo {author} {\bibfnamefont {N.}~\bibnamefont {Ishii}}, \ and\ \bibinfo
  {author} {\bibfnamefont {A.~W.}\ \bibnamefont {Thomas}},\ }\href {\doibase
  10.1016/S0375-9474(03)00635-3} {\bibfield  {journal} {\bibinfo  {journal}
  {Nucl. Phys.}\ }\textbf {\bibinfo {volume} {A720}},\ \bibinfo {pages} {95}
  (\bibinfo {year} {2003})},\ \Eprint {http://arxiv.org/abs/nucl-th/0210067}
  {arXiv:nucl-th/0210067 [nucl-th]} \BibitemShut {NoStop}%
\bibitem [{\citenamefont {Lawley}\ \emph {et~al.}(2006)\citenamefont {Lawley},
  \citenamefont {Bentz},\ and\ \citenamefont {Thomas}}]{Lawley:2005ru}%
  \BibitemOpen
  \bibfield  {author} {\bibinfo {author} {\bibfnamefont {S.}~\bibnamefont
  {Lawley}}, \bibinfo {author} {\bibfnamefont {W.}~\bibnamefont {Bentz}}, \
  and\ \bibinfo {author} {\bibfnamefont {A.~W.}\ \bibnamefont {Thomas}},\
  }\href {\doibase 10.1016/j.physletb.2005.11.025} {\bibfield  {journal}
  {\bibinfo  {journal} {Phys. Lett.}\ }\textbf {\bibinfo {volume} {B632}},\
  \bibinfo {pages} {495} (\bibinfo {year} {2006})},\ \Eprint
  {http://arxiv.org/abs/nucl-th/0504020} {arXiv:nucl-th/0504020 [nucl-th]}
  \BibitemShut {NoStop}%
\bibitem [{\citenamefont {Kiriyama}\ \emph {et~al.}(2001)\citenamefont
  {Kiriyama}, \citenamefont {Yasui},\ and\ \citenamefont
  {Toki}}]{Kiriyama:2001ud}%
  \BibitemOpen
  \bibfield  {author} {\bibinfo {author} {\bibfnamefont {O.}~\bibnamefont
  {Kiriyama}}, \bibinfo {author} {\bibfnamefont {S.}~\bibnamefont {Yasui}}, \
  and\ \bibinfo {author} {\bibfnamefont {H.}~\bibnamefont {Toki}},\ }\href
  {\doibase 10.1142/S0218301301000642} {\bibfield  {journal} {\bibinfo
  {journal} {Int. J. Mod. Phys.}\ }\textbf {\bibinfo {volume} {E10}},\ \bibinfo
  {pages} {501} (\bibinfo {year} {2001})},\ \Eprint
  {http://arxiv.org/abs/hep-ph/0105170} {arXiv:hep-ph/0105170 [hep-ph]}
  \BibitemShut {NoStop}%
\bibitem [{\citenamefont {Huang}\ and\ \citenamefont
  {Shovkovy}(2003)}]{Huang:2003xd}%
  \BibitemOpen
  \bibfield  {author} {\bibinfo {author} {\bibfnamefont {M.}~\bibnamefont
  {Huang}}\ and\ \bibinfo {author} {\bibfnamefont {I.}~\bibnamefont
  {Shovkovy}},\ }\href {\doibase 10.1016/j.nuclphysa.2003.10.005} {\bibfield
  {journal} {\bibinfo  {journal} {Nucl. Phys.}\ }\textbf {\bibinfo {volume}
  {A729}},\ \bibinfo {pages} {835} (\bibinfo {year} {2003})},\ \Eprint
  {http://arxiv.org/abs/hep-ph/0307273} {arXiv:hep-ph/0307273 [hep-ph]}
  \BibitemShut {NoStop}%
\bibitem [{\citenamefont {Aubert}\ \emph {et~al.}(1983)\citenamefont {Aubert}
  \emph {et~al.}}]{Aubert:1983xm}%
  \BibitemOpen
  \bibfield  {author} {\bibinfo {author} {\bibfnamefont {J.~J.}\ \bibnamefont
  {Aubert}} \emph {et~al.} (\bibinfo {collaboration} {European Muon}),\ }\href
  {\doibase 10.1016/0370-2693(83)90437-9} {\bibfield  {journal} {\bibinfo
  {journal} {Phys. Lett.}\ }\textbf {\bibinfo {volume} {123B}},\ \bibinfo
  {pages} {275} (\bibinfo {year} {1983})}\BibitemShut {NoStop}%
\bibitem [{\citenamefont {Geesaman}\ \emph {et~al.}(1995)\citenamefont
  {Geesaman}, \citenamefont {Saito},\ and\ \citenamefont
  {Thomas}}]{Geesaman:1995yd}%
  \BibitemOpen
  \bibfield  {author} {\bibinfo {author} {\bibfnamefont {D.~F.}\ \bibnamefont
  {Geesaman}}, \bibinfo {author} {\bibfnamefont {K.}~\bibnamefont {Saito}}, \
  and\ \bibinfo {author} {\bibfnamefont {A.~W.}\ \bibnamefont {Thomas}},\
  }\href {\doibase 10.1146/annurev.ns.45.120195.002005} {\bibfield  {journal}
  {\bibinfo  {journal} {Ann. Rev. Nucl. Part. Sci.}\ }\textbf {\bibinfo
  {volume} {45}},\ \bibinfo {pages} {337} (\bibinfo {year} {1995})}\BibitemShut
  {NoStop}%
\bibitem [{\citenamefont {Malace}\ \emph {et~al.}(2014)\citenamefont {Malace},
  \citenamefont {Gaskell}, \citenamefont {Higinbotham},\ and\ \citenamefont
  {Cloët}}]{Malace:2014uea}%
  \BibitemOpen
  \bibfield  {author} {\bibinfo {author} {\bibfnamefont {S.}~\bibnamefont
  {Malace}}, \bibinfo {author} {\bibfnamefont {D.}~\bibnamefont {Gaskell}},
  \bibinfo {author} {\bibfnamefont {D.~W.}\ \bibnamefont {Higinbotham}}, \ and\
  \bibinfo {author} {\bibfnamefont {I.}~\bibnamefont {Cloët}},\ }\href
  {\doibase 10.1142/S0218301314300136} {\bibfield  {journal} {\bibinfo
  {journal} {Int. J. Mod. Phys.}\ }\textbf {\bibinfo {volume} {E23}},\ \bibinfo
  {pages} {1430013} (\bibinfo {year} {2014})},\ \Eprint
  {http://arxiv.org/abs/1405.1270} {arXiv:1405.1270 [nucl-ex]} \BibitemShut
  {NoStop}%
\bibitem [{\citenamefont {Cloët}\ \emph {et~al.}(2019)\citenamefont {Cloët}
  \emph {et~al.}}]{Cloet:2019mql}%
  \BibitemOpen
  \bibfield  {author} {\bibinfo {author} {\bibfnamefont {I.~C.}\ \bibnamefont
  {Cloët}} \emph {et~al.},\ }\href {\doibase 10.1088/1361-6471/ab2731}
  {\bibfield  {journal} {\bibinfo  {journal} {J. Phys.}\ }\textbf {\bibinfo
  {volume} {G46}},\ \bibinfo {pages} {093001} (\bibinfo {year} {2019})},\
  \Eprint {http://arxiv.org/abs/1902.10572} {arXiv:1902.10572 [nucl-ex]}
  \BibitemShut {NoStop}%
\bibitem [{\citenamefont {Detmold}\ \emph {et~al.}(2006)\citenamefont
  {Detmold}, \citenamefont {Miller},\ and\ \citenamefont
  {Smith}}]{Detmold:2005cb}%
  \BibitemOpen
  \bibfield  {author} {\bibinfo {author} {\bibfnamefont {W.}~\bibnamefont
  {Detmold}}, \bibinfo {author} {\bibfnamefont {G.~A.}\ \bibnamefont {Miller}},
  \ and\ \bibinfo {author} {\bibfnamefont {J.~R.}\ \bibnamefont {Smith}},\
  }\href {\doibase 10.1103/PhysRevC.73.015204} {\bibfield  {journal} {\bibinfo
  {journal} {Phys. Rev.}\ }\textbf {\bibinfo {volume} {C73}},\ \bibinfo {pages}
  {015204} (\bibinfo {year} {2006})},\ \Eprint
  {http://arxiv.org/abs/nucl-th/0509033} {arXiv:nucl-th/0509033 [nucl-th]}
  \BibitemShut {NoStop}%
\bibitem [{\citenamefont {Tolman}(1939)}]{Tolman:1939jz}%
  \BibitemOpen
  \bibfield  {author} {\bibinfo {author} {\bibfnamefont {R.~C.}\ \bibnamefont
  {Tolman}},\ }\href {\doibase 10.1103/PhysRev.55.364} {\bibfield  {journal}
  {\bibinfo  {journal} {Phys. Rev.}\ }\textbf {\bibinfo {volume} {55}},\
  \bibinfo {pages} {364} (\bibinfo {year} {1939})}\BibitemShut {NoStop}%
\bibitem [{\citenamefont {Oppenheimer}\ and\ \citenamefont
  {Volkoff}(1939)}]{Oppenheimer:1939ne}%
  \BibitemOpen
  \bibfield  {author} {\bibinfo {author} {\bibfnamefont {J.~R.}\ \bibnamefont
  {Oppenheimer}}\ and\ \bibinfo {author} {\bibfnamefont {G.~M.}\ \bibnamefont
  {Volkoff}},\ }\href {\doibase 10.1103/PhysRev.55.374} {\bibfield  {journal}
  {\bibinfo  {journal} {Phys. Rev.}\ }\textbf {\bibinfo {volume} {55}},\
  \bibinfo {pages} {374} (\bibinfo {year} {1939})}\BibitemShut {NoStop}%
\end{thebibliography}

%

\end{document}